\documentclass[12pt]{article}
\usepackage{graphicx}
\newcommand{\grad}{\mbox{\boldmath{$\nabla$}}}

\newcommand{\bb}{\mbox{\boldmath{$B$}}_0}

\newcommand{\bxi}{\mbox{\boldmath{$\xi$}}}
\newcommand{\boldeta}{\mbox{\boldmath{$\eta$}}}

\newcommand{\zz}{\bf\hat{z}}
\newcommand{\xx}{\bf\hat{x}}
\newcommand{\yy}{\bf\hat{y}}

\begin{document}
\begin{center}
{\large \bf Explosive Instability and Erupting Flux Tubes in a Magnetised Plasma Atmosphere}
\\[3mm]
{\bf  S C Cowley$^{1,2}$, B Cowley$^3$, S A Henneberg$^4$ and H R Wilson$^4$}
\end{center}
{\it  $^1$ CCFE, Culham Science Centre, Abingdon, Oxon.  OX14 3DB, UK.
\\
$^2$ Department of Physics, Imperial College, Prince Consort Road, London SW7 2BZ UK 
\\
$^3$Department of Physics, University College London, London, UK
\\
$^4$York Plasma Institute and Department of Physics, University of York, Heslington, York YO10 5DD UK}

\abstract
The eruption of multiple flux tubes in a magnetised plasma atmosphere is proposed as a mechanism for explosive release of energy in plasmas.  Linearly stable isolated flux tubes are shown to be metastable in a box model magnetised atmosphere in which ends of the field lines are embedded in conducting walls.  The energy released by destabilising such field lines can be a significant fraction of the gravitational energy stored in the system.  This energy can be released in a fast dynamical time.

\section{Introduction}
\label{intro}
The explosive release of energy from magnetically confined plasmas produces some of the most dramatic and destructive natural phenomena.  In such events a slowly evolving plasma suddenly erupts  releasing a significant fraction of its stored magnetic, gravitational or pressure energy in a few tens of dynamical times (which is typically the Alfv\'{e}n time or the free fall time).  The stored energy is converted into some combination of heat, energetic particles, fast plasma flows and/or radiation.  Tokamak disruptions \cite{iter}, solar flares \cite{shibata}, coronal mass ejections \cite{chen}, magnetospheric substorms \cite{Pulkkinen} and edge localised modes in tokamaks  \cite{zohm, connor, suttrop, iaea, kirk, kirk2, Eich, smallELM}   all exhibit this type of explosive behaviour.   While there are a number of theories of these phenomena many of the central questions are still without quantitative answers.  For example.  What triggers the instability?  What sets the timescale?  How much energy is released?  How much energy is converted into energetic particles?  Are there universal mechanisms?  There is probably more than one mechanism.  Perhaps there are no universal mechanisms for these phenomena but certainly they share common issues.  Here we propose an explosive scenario where multiple metastable flux tubes are destabilised and erupt on Alfv\'{e}n timescales.  This scenario is an extension of earlier weakly nonlinear analyses  \cite{cowart, cowartbright, HFC,  fong, fong2, wilcow, CWHF, HFCCKP}.  These papers demonstrate that just above their linear stability threshold {\em all} fine scale pressure and gravitational instabilities obey a generic equation that yields explosive dynamics.  In this paper (except for Appendix~(\ref{general})) we address the gravitational stability and fully nonlinear flux tube dynamics of a simple slab atmosphere that is line tied ({\em i.e.} one in which the ends of the field lines are embedded in conducting walls and therefore immovable).  We suspect that the flux tube picture that emerges from our simple analysis captures some generic elements of the fully nonlinear evolution of {\em all} fine scale gravitational and pressure driven instabilities ({\em e.g.} ballooning modes  \cite{kulsrud, furth, CHT, hood}).  Although the proposed scenario superficially resembles some observations of edge localised modes \cite{kirk2} and magnetospheric instability \cite{HFCCKP} further work is needed to establish any quantitative relationship to data.  

The dynamics of flux tubes in plasmas is an old subject -- we cannot do justice to it all here.  The stability of plasmas to the interchange of flux tubes was, we believe, first discussed by Edward Teller in 1954 in a classified meeting at Princeton on magnetically confined plasmas for fusion -- see discussion in Chapter 9. of \cite{bishop}.  This led to the classic analysis of Rosenbluth and Longmire in \cite{longmire} (see also \cite{bernstein}).  Parker made extensive use of a circular flux tube approximation in discussing dynamics in the Solar convection zone; work that is reviewed in his famous monograph {\em Cosmical Magnetic Fields} \cite{Parker}.  Spruit developed the flux tube theory including examining the stability of horizontal magnetic field in an atmosphere \cite{spruit}. In fusion research the theory of 'blobs' (isolated field aligned plasma structures that are similar to flux tubes but not necessarily frozen to the field) has been extensively researched \cite{Ippolito}.  Fan summarises the work on gravitational stability and flux tubes for solar convection in \cite{Fan1} see also \cite{Fan2}, \cite{hughes} and \cite{proctor}.  Much of the research on flux tubes concerns tubes in an otherwise unmagnetised plasma where circular tubes might be expected.  The discussion here emphasises for the first time that tubes are expected to be highly elliptical when passing through a magnetised plasma.  This shape minimises the stabilising sideways perturbations of the surrounding field.  It is also the first time that the metastability properties have been calculated and the nonlinear consequences explored.

To motivate our development let us examine a very familiar zero dimensional bifurcation problem that captures {\em some} of the features of the eruption of a single flux tube.  Consider a normalised (non-dimensional) displacement/amplitude of the eruption $A(t)$.  Let the system pass slowly through marginal stability (at $t=0$) with a long timescale $\tau_E$.  We denote the typical growth rate of an instability (when it is well above the marginal point) to be $\gamma_A$ -- an Alfv\'{e}nic rate.  For the systems of interest $\gamma_A\tau_E \gg 1$.  Let the normalised "potential energy" of the system be given by the function $V(\frac{t}{\tau_E}, A)$.  We expect that at the point of triggering the system must be close to a linear instability boundary since otherwise a large perturbation is needed to trigger eruption.  For $A\ll 1$ and close to the linear stability boundary, $|t|\ll |\tau_E|$,we expand:
\begin{equation}
V(\frac{t}{\tau_E}, A) = - \gamma_A^2\left(\frac{t}{2\tau_E} A^2 + \frac{1}{3}k_1  A^3 +  \frac{1}{4}k_2  A^4\; ......\right)
\label{potential}\end{equation}
where $k_1$ and $k_2$ are constants of order one.  Then the equation of motion is:
\begin{equation}
\frac{d^2 A}{d t^2} = -\frac{\partial V}{\partial A} \sim \gamma_A^2\left(\frac{t}{\tau_E} A + k_1  A ^2 +  k_2  A^3\; ......\right).
\label{bifurc}\end{equation}
The linear motion obeys Airy's equation so that for $t\gg (\tau_E\gamma_A^{-2})^{1/3}$ the amplitude is $A\sim A_0\exp{(\frac{2\gamma_A t^{3/2}}{3\sqrt{\tau_E}})}$ where $A_0$ is a constant typically of order the initial amplitude.  Thus the linear system does not reach an Alfv\'{e}nic growth rate until $t\sim \tau_E$ by which time it has exponentiated by a factor $e^{\frac{2}{3}\gamma_A\tau_E}$.  Given that in many systems $\gamma_A\tau_E >100$ it is likely that the initial perturbation due to noise  is sufficiently large ($A_0\gg e^{-\frac{2}{3}\gamma_A\tau_E}$) that the system is nonlinear long before it reaches an Alfv\'{e}nic linear growth rate.  Then nonlinear dynamics begins while $A\ll 1$ and $t\ll \tau_E$.

\begin{figure}[h]
\setlength{\unitlength}{1cm}
\begin{center}
{\includegraphics[angle=0, width=14.0cm, totalheight=9.0cm,trim=0 0 0 0,clip]{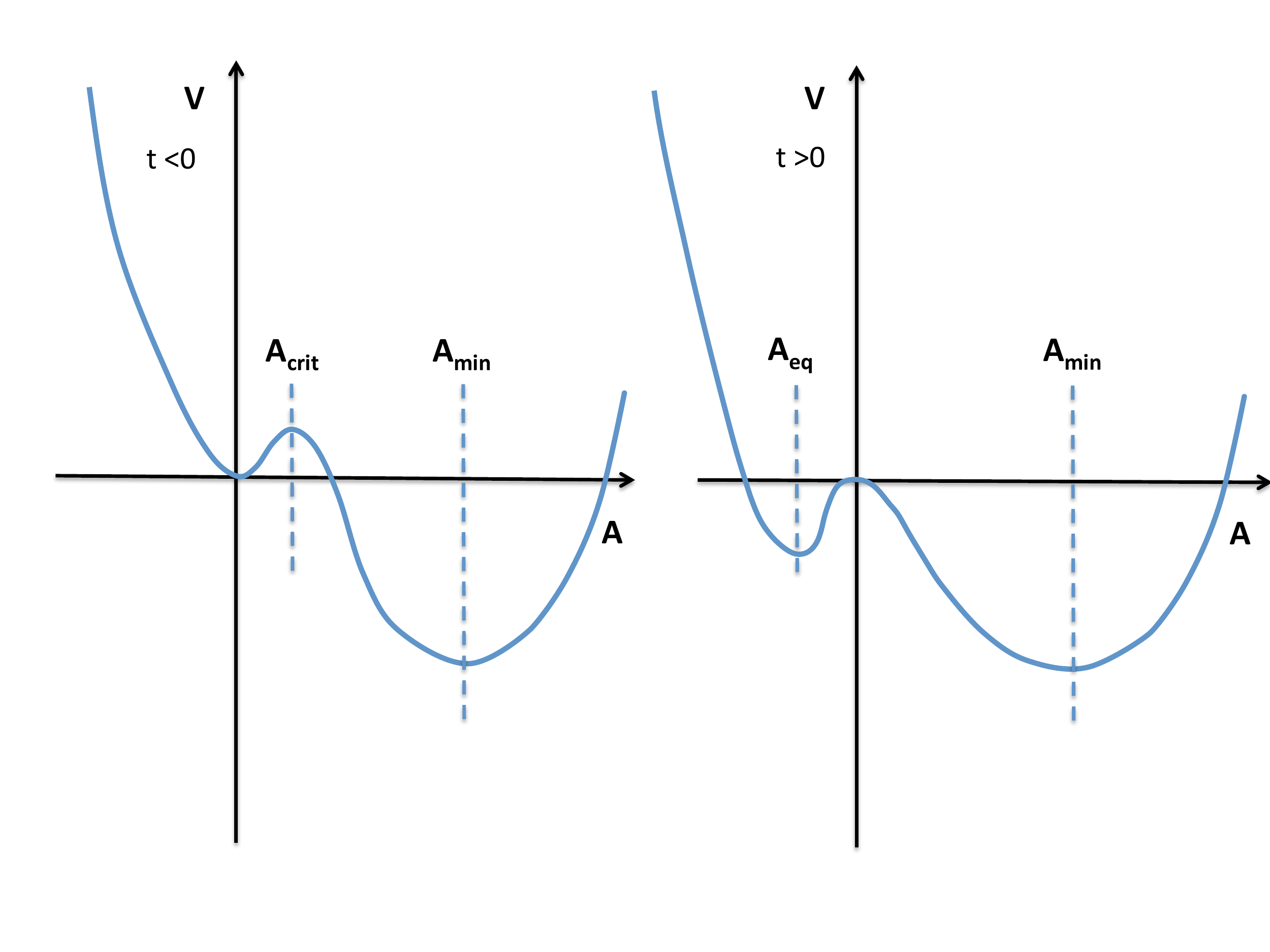}
}
\vskip -0.3 truein
\caption{\textit{Sketch of a model transcritical bifurcation potential with $k_1>0$ and $t\ll \tau_E$. On the left is the metastable ($t<0$) case where supercritical positive perturbations with $A> A_{crit} = -\frac{t}{k_1\tau_E} > 0$ grow explosively and subcritical perturbations with  $A< A_{crit}$ are stable.  With dissipation the supercritical perturbation will settle into the minimum energy $A=A_{min}$ state (usually beyond the validity of the expansion in Eqs.~(\ref{potential}) and (\ref{bifurc})).  On the right 
is the linearly unstable case where for $A>0$ dissipative explosive motion will take the system to the energy minimum $A=A_{min}$ and for $A<0$ dissipative motion will take the system to the local minimum $A=A_{eq}$.  The potential energy released by going to $A_{min}$ is finite (${\cal O}(\gamma_A^2)$) whereas the energy released going to $A_{eq}$ is small close to marginal stability -- {\em i.e.} when $t\ll\tau_E$, $V(\frac{t}{\tau_E}, A_{eq})\sim-\frac{\gamma_A^2}{6k_1^2}(\frac{t}{\tau_E})^3 \ll \gamma_A^2$. }} \end{center}\label{potpic}\end{figure}

In many systems  symmetry requires that $k_1 =0$.  However in the models developed here there is no such constraint (see \cite{fong2}).  Thus for small $A$ we may ignore the $k_2$ term in Eq.~(\ref{bifurc}) and we have a system with a form of {\em transcritical bifurcation} \cite{strog}.   We shall take $k_1>0$ so that the explosive drive/force in Eq.~(\ref{bifurc}) is in the positive direction.  In Figure~(\ref{potpic}) we illustrate a typical $V$ for just below ($t<0$) and just above ($t>0$) marginal stability.  When the nonlinear term dominates and A is positive, $\frac{t}{\tau_E}  \ll k_1  A $, the explosive solution is:
\begin{equation}
A\sim \frac{6}{k_1\gamma_A^2(t_0 - t)^2}.
\label{exploder}\end{equation}
If $A(t=0) \gg (\gamma_A\tau_E)^{-2/3}/k_1$, then the dynamics is always nonlinear and $\gamma_A t_0 \sim \sqrt{\frac{6}{k_1 A(t=0)}}$.  Otherwise there is a linear phase
before the nonlinear explosive phase begins.  Clearly with the explosive nonlinearity the instantaneous growth rate $\frac{d\log{A}}{dt} \sim \gamma_A\sqrt{\frac{k_1 A}{6}}$ always approaches Alfv\'{e}nic  values as $A$ becomes finite.   The dynamics does not, of course, become singular (as it would if Eq.~(\ref{exploder}) continued to hold for all displacement amplitudes) since the expansion of the potential in powers of $A$ must
break down.  With dissipation the system must seek out the minimum of the potential energy $V(\frac{t}{\tau_E}, A_{min})$ where $\frac{\partial V}{\partial A}|_{A=A_{min}} =0$ -- see Figure~(\ref{potpic}).  When $k_1\neq 0$ and $0< t \ll |\tau_E|$ there are at least two minima: a local minimum for $A= A_{eq} \sim - t/(k_1\tau_E) <0$ (a negative displacement) but this does not yield much energy release (Figure~(\ref{potpic})) and at least one minimum at finite positive A, $A=A_{min}$, which can yield a finite energy release.  Note that around this finite minimum $A=A_{min}$, $V$ is not usually well modelled by the Taylor series of Eqs.~(\ref{potential}) and (\ref{bifurc}).  We illustrate one finite minimum for simplicity in Figure~(\ref{potpic}).   Without dissipation the system will execute some slowly evolving nonlinear oscillation in which the action is conserved.  With viscous type dissipation, for example a drag term $\nu\frac{dA}{dt}$ on the right had side of Eq.~(\ref{bifurc}), the motion initiated by a small random noise source will predominantly settle down to the stationary state at $A=A_{min}$ in preference to the $A=A_{eq}$ state.  The energy that is dissipated (as heat etc.) will be equal to $V(0, 0) - V(0, A_{min})$, where we have dropped corrections of order $\frac{t}{\tau_E}$.  We note again that {\it transcritical bifurcations} lead to asymmetric response -- e.g. in our case with $k_1>0$ explosive motion in the positive direction only.

Ideal Magnetohydrodynamic (MHD) instabilities just above the marginal stability point  are of two distinct types: either {\em global} or {\em local} instabilities.  The {\em kink mode} driven by current in a plasma cylinder is the archetype of the {\em global} (finite scale) MHD instability.  At a critical current the plasma becomes unstable to a single helical kink mode (see for example \cite{Freidberg}).  Early studies \cite{Fredricks, Rutherford} on simple cylindrical cases showed that crossing the linear stability boundary resulted in bifurcation to nearby helical equilibria.  Such helical states have also been predicted \cite{white} and observed \cite{padova} to be the result of crossing a global resistive instability boundary.  Marginally stable gravitational and pressure driven ideal MHD instabilities ({\em e.g.} the ideal ballooning mode \cite{kulsrud, furth, CHT}) are local instabilities with an infinitesimal scale perpendicular to the field lines.  The simple line tied equilibria addressed in this paper passes through the marginal linear instability threshold in this way (see Appendix~(\ref{threshold})).  In such transitions an infinity of modes (all with infinitesimal perpendicular scale) becomes unstable just above marginal stability and the threshold dynamics is complex.  A previous series of papers \cite{cowart, cowartbright, HFC, fong, fong2, wilcow, CWHF} examined the weakly nonlinear evolution of gravitational and pressure driven instabilities.  These instabilities exhibit generic explosive dynamics when passing slowly through the linear instability boundary.  Narrowing fingers of plasma accelerate and push aside the surrounding magnetic field to release energy -- see Section (\ref{pert}).  It is not clear from the weakly nonlinear analysis what happens to these fingers.  Simulations   confirmed the nonlinear instability \cite{myers}and the formation of narrow fingers \cite{myers,fong2, Zhu} but lose resolution before the instability reaches saturation.  Here we propose that the fingers evolve into eruptions of multiple elliptical flux tubes.  We derive equations of motion for an isolated elliptical tube in a simple one dimensional line tied equilibrium (with gravity) (see Section~(\ref{fluxtube})).   This simple one dimensional equilibrium illustrates generic behaviour.  The tubes have dynamics that is analogous to the simple zero dimensional {\em transcritical} bifurcation dynamics described above.  Specifically they have nonlinear instability drive (with a quadratic amplitude dependence of the force) so that even linearly stable tubes are metastable and will erupt given sufficient perturbation.  Thus triggering metastable flux tubes by either a finite  perturbation or by linear instability can yield an explosive release of energy on an Alfv\'{e}n timescale.  Erupting metastable flux tubes evolve to stable three dimensional equilibria with finite displacements.  The energy released by each tube in the evolution to these stable states is a finite fraction of the energy stored in the tube.  The existence of a minimum energy state for the flux tube is not however guaranteed.  Indeed, in some cases with relatively weak initial magnetic field, the flux tubes erupt to a singular state with zero magnetic field (flux expulsion) along part of the tube,  see Section~(\ref{Expulsion}).  We do not know how such singular states evolve or if such eruptions occur in nature.  

The analysis of the one dimensional dynamics of isolated elliptical flux tubes as developed in Section~(\ref{fluxtube}) misses the crucial issue of how the tubes interact.  Specifically it does not address how energy released by the eruption of one tube might be used to destabilise other metastable tubes.  Nor does it determine the fraction of the metastable tubes that erupt.  The weakly nonlinear case demonstrated that the perturbation spread from the linearly unstable region to the metastable region -- destabilising the metastable field lines, \cite{cowart, cowartbright, HFC, wilcow, CWHF, fong}.  This process was called {\em detonation} \cite{cowart} because of the partial analogy with chemical explosives.   It is not clear how detonation proceeds.  However, in the model equilibrium in Subsection~(\ref{model}) the energy needed to destabilise  all the metastable field lines is considerably less than the energy released by these field lines (see Figures~(\ref{E1sat/crit3}, \ref{C2Esat})).  Thus it seems plausible that with a small fraction of field lines close to marginal stability a modest perturbation will destabilise the nearly marginal field lines.  Then the energy released by the motion of these field lines will progressively destabilise a significant fraction of the metastable field lines.  Estimates show that a significant fraction of all the tubes can be destabilised so that a considerable fraction of the stored energy can be released in a few Alfv\'{e}n times by this eruptive scenario (see Section~(\ref{discussion})).  Triggering happens when the system gets close enough to marginal linear stability that noise is sufficiently large to destabilise nearly marginally unstable field lines.  We emphasise that while our treatment of both the weakly nonlinear evolution and the dynamics of isolated flux tubes is rigorous asymptotic analysis the discussion of the evolution of multiple flux tubes is speculative.

In the next section, Section~(\ref{mod}), we define the one dimensional line tied equilibrium and the equations to be solved. The weakly nonlinear case is summarised in
Section~(\ref{pert}) -- this is a small extension of the work in \cite{cowart}.  The theory of isolated elliptical flux tube dynamics is developed in Section~(\ref{fluxtube}). In Subsection~(\ref{small amp}) we show how the small amplitude behaviour of isolated flux tubes connects to the weakly nonlinear results.  The singular case, when the field in the erupting field lines becomes zero, is treated in Subsection~(\ref{Expulsion}).  In Subsection~(\ref{model}) we solve the equations for flux tube motion in a simple model atmosphere.  Two specific cases are presented in Subsection~(\ref{numerical}): CASE 1. illustrates a case with no flux expelled erupted states and CASE 2. with a region of erupted field lines that are in a flux expelled state.  We choose parameters so that we change the field strength (and therefore the plasma $\beta$) keeping the density profile and $B/L$ constant (where $L$ is the length of the field lines).  Then both cases have the same growth rate and quadratic nonlinearity profile that is just above marginal stability in a very small section of the profile ($1.0945< {\tilde x}_0 < 1.1288$ where ${\tilde x}_0 $ is the {\em initial} height of the field line).  The saturation of the eruptions in each case is, however, different.  In Case 1. the saturated equilibrium state of the field lines from $0.23<{\tilde x}_0 <1.2$ has a lower energy than the initial state -- see Figure~(\ref{C1sat1}).  Those field lines from  $0.23<{\tilde x}_0 <1.0945$ and $1.1288<{\tilde x}_0 <1.2$ are metastable and must be displaced finitely to be destabilised.   The energy released by metastable field lines and the critical energy to excite the field line is shown for Case 1. in Figure~(\ref{E1sat/crit3}).  The lower field lines overtake the upper lines and the maximum energy release is associated with metastable field lines (not linearly unstable field lines).  The field strength in Case 2. is about $5\%$ less than in Case 1.  In this case all field lines between $0.13<{\tilde x}_0 <1.2$ minimise their energy in a saturated displaced equilibrium.  The lines between ${\tilde x} \sim 0.295$ to ${\tilde x} \sim 0.62$ evolve to the singular flux expelled state. The energy released by metastable field lines and the critical energy to excite the field line is shown in Figure~(\ref{C2Esat}).  In Section(\ref{discussion}) we discuss qualitatively the shape of the flux tubes, the number of destabilised flux tubes, their mutual interaction and present our conclusions.  The proof that in the line tied slab the unstable modes just above the linear stability threshold have infinitesimal perpendicular scales is given in Appendix~(\ref{threshold}).  In Appendix~(\ref{general}) we derive the equations governing elliptical flux tube motion in general three dimensional magneto-static equilibria.  These equations have the same structure as the one dimensional case but may be useful for analysis of more realistic cases.  Finally in Appendix~(\ref{BoundaryLayer}) we investigate the structure of the boundary layer that develops at the line tied walls -- this is a complicating issue that does not affect the results of the main body of the paper.

\section{Equilibrium and Equations}
\label{mod}
In this section we outline the one dimensional model geometry and the equations to be solved.  
\subsection{Equilibrium}
\label{mod}
We consider a simple one dimensional line tied magnetised atmosphere with magnetic field ($\bb$), gravitational acceleration ($\bf g$), pressure ($p_0$) and density ($\rho_0$) given by:
\begin{equation}
{\bb} = B_0(x){\zz},\;\;\;\; {\bf g} = -g{\xx},\;\;\;\;\; p_0 = p_0(x) \;\;\; and \;\;\; \rho_0 = \rho_0(x).
\label{defin}\end{equation}
The equilibrium force balance is: 
\begin{equation}
\frac{d}{dx}\left[ \frac{B_0^2}{2} + p_0 \right] = -g\rho_0.
\label{equilib}\end{equation}
Thus to fully specify the equilibrium we must give two of the three functions of height, $B_0(x)$, $p_0(x)$ or $\rho_0(x)$ (and a boundary condition if $\rho_0$ is one of the two).  Note we use magnetic field units where $B^2$ has units of energy density.
\subsection{Perturbation: equations and boundary conditions}
\label{mod}
 We adopt a simple ideal MHD system with scalar viscosity that captures the essential dynamics of fast explosive motion driven by the gravitational potential energy.  The ideal motion drives the system to small scales thus dissipative processes are inevitably important.  These processes are often not well modelled by a scalar viscosity -- nonetheless the 
basic picture of instability and metastability is, we believe, a robust generic feature of magnetically confined systems.   The equation of motion is:
\begin{equation}
\rho \left(\frac{\partial {\bf v}}{\partial t} + {\bf v}\cdot\nabla{\bf v}\right) = -\nabla\left( p + \frac{B^2}{2}\right) +{\bf B}\cdot\nabla{\bf B} -\rho g\xx +  \nu\rho\nabla^2{\bf v}
\label{moment}\end{equation}
where $\nu$ is the small viscosity coefficient.  The magnetic field obeys the familiar equation for frozen in field:
\begin{equation}
\frac{\partial {\bf B}}{\partial t}  = \nabla\times( {\bf v}\times{\bf B}).
\label{magnetic}\end{equation}
{\em i.e.} ${\bf E} = -  {\bf v}\times{\bf B}$.  Density obeys the continuity equation:
\begin{equation}
\frac{\partial {\rho}}{\partial t}  = -\nabla\cdot( {\bf v}\rho).
\label{density}\end{equation}
The field lines are tied to walls at $z=0$ and $z=L$.  At the boundaries $z=0$ and $z=L$
the $x$ and $y$ components of the Electric field are set to zero -- {\em i.e.} $E_x(z=0, x, y, t)= E_x(z=L, x, y, t)= E_y(z=0, x, y, t) = E_y(z=L, x, y, t) = 0$. We also set the pressure and density to be unperturbed at the boundaries -- {\em i.e.} $p(z=0, x, y, t) = p(z=L, x, y, t) = p_0(x)$ and $\rho(z=0, x, y, t) = \rho(z=L, x, y, t) = \rho_0(x)$; thus motion along the field through the boundary is allowed ({\em i.e.} ${\bf v\cdot B}\neq 0$ at $z=0,L$).  These boundary conditions differ from those in \cite{cowart} where all flow at the boundary is zero.  The boundary conditions adopted in this paper are closely related to the behaviour on infinite field lines where the perturbed field line displacement, pressure and density must vanish as distance along the field line goes to infinity.  

We will take the thermal conduction along the field line to be fast so that the temperature is constant on the field line {\em i.e.} $T(x,y,z,t) = T_0(x_0)$ where $x_0$ is the height of the field line at the walls (at $z=0,L$).  Note that since the field is frozen into the stationary wall and the moving plasma $x_0$ is the original (Lagrangian) height of the field line.  The pressure is then obtained from $p(x,y,z,t) = \rho (x,y,z,t)T_0(x_0)/m$ where $m$ is the ion mass.  This is equivalent to the usual adiabatic equation for pressure with the ratio of specific heats equal to one ({\em i.e.} $p$ obeys the same equation as $\rho$, Eq.~(\ref{density})).

The slab system with the line tied boundary conditions (and no viscosity) will become linearly unstable above a critical length -- see below.  The passage through marginal stability may be effected by either lengthening the box or evolution of the (density, pressure or magnetic field) profiles by diffusion, heating or cooling.  The first modes to become unstable have $k_y\rightarrow\infty$ where $k_y$ is the wavenumber in the $y$ direction -- a proof is given in Appendix \ref{threshold}.  This proof is a simple extension for our chosen boundary conditions of the proof in \cite{Zweibel} which used elements of \cite{newcomb} and \cite{gilman}.

\section{Small Amplitude Nonlinear Motion.}
\label{pert}

In this section we calculate the early, small amplitude, nonlinear evolution of fine $y$ scale perturbations when the system is just above the linear marginal stability threshold.  The treatment closely follows the development of Cowley and Artun in 1997 \cite{cowart} -- thus we omit considerable detail.  We measure the distance above marginal stability by a dummy large parameter $n$ where the growth rate of the most unstable perturbation is order $n^{-1/2}\sqrt{g\rho'/\rho}$.  The typical y and z wave numbers are $k_y\sim {\cal O}(n/L),\;\;\; k_z\sim {\cal O}(1/L)$ and the mode is localised in x over a distance $\Delta x\sim {\cal O}(n^{-1/2}L)$.  The viscosity is treated as small $\nu \sim {\cal O}(n^{-5/2})$. The Lagrangian displacement of the plasma, $\bxi$ is of the form:
\begin{equation}
{\bxi} = \frac{1}{n}\left[\xi_x {\xx}  + \xi_z {\zz} \right] + {\frac{1}{n^{3/2}}} \left[\xi_y\yy\right].
\end{equation}
The perpendicular motion of field lines is predominately in the x direction and the structure is elongated in the x direction compared to the y direction. The y motion is small.  This structure maximises the motion in the direction of gravity thereby enhancing the release of potential energy. Expansion of the MHD equations in powers of $n^{-1/2}$ first yields the form of displacement:
\begin{equation}
\xi_x = \xi (x_0, y_0, t)\sin{(\frac{\pi z}{L})},\;\;\;\;\; \xi_z = -\left(\frac{\rho_0(x_0)g}{p_0(x_0)}\right)\xi (x_0, y_0, t)\cos{(\frac{\pi z}{L})}
\end{equation}
and perpendicular incompressibility $\frac{\partial \xi_y}{\partial y_0} + \frac{\partial \xi_x}{\partial x_0}=0$.  The Cowley and Artun \cite{cowart} paper was formulated entirely in Lagrangian variables where the current position of a plasma element, $\bf r$, is related to the initial position of the same element, $\bf r_0$, by $\bf r = r_0 + \bxi$.  Here we have expressed the displacement in mixed Lagrangian-Eulerian variables -- z is the current (Eulerian) z position of a plasma element and $x_0$ and $y_0$ the initial (Lagrangian) x and y positions of the plasma element.  This rather complicated representation is convenient since it makes the lowest order displacement satisfy exactly the boundary conditions at $z=0,L$.  In higher order we obtain the evolution equation for $\xi$:
\begin{equation}
C_0\frac{\partial^2\xi}{\partial t^2} = \Gamma^2(x_0)\xi + C_2\frac{\partial^2 u}{\partial x_0^2} + C_3\xi\frac{\partial^2{\overline {\xi^2}}}{\partial x_0^2} + C_4 (\xi^2 - {\bar \xi^2})  + \nu\frac{\partial^2}{\partial y_0^2}(\frac{\partial\xi}{\partial t})
\label{nonlineareq}\end{equation}
where ${\overline {\xi^2}}$ is the $y_0$ average of the squared displacement, $\xi^2$ and $\frac{\partial^2 u}{\partial y_0^2} = \xi$.  The local linear growth rate $\Gamma_0$ is given by:
\begin{eqnarray}
\Gamma_0^2(x_0) = -\frac{B_0^2 \pi^2}{\rho_0 L^2} +\frac{\rho_0g^2}{p_0} + \frac{g}{\rho_0}\frac{d\rho_0}{dx_0},
\label{Gamma}\end{eqnarray}
where the first term is the stabilising {\em field line bending}, the second term is sometimes called the {\em Parker instability drive} and the third term is the {\em Rayleigh-Taylor instability drive}.  Note that we have assumed that $\Gamma_0 \sim n^{-1/2}\sqrt{g\rho'/\rho}$ so that the terms in Eq.~(\ref{Gamma}) cancel to dominant order.  Using the equilibrium relation, Eq.~(\ref{equilib}) we can write $\Gamma_0^2(x_0)$ in a form perhaps more familiar to some readers:
\begin{eqnarray}
\Gamma_0^2(x_0) = -\frac{B_0^2 \pi^2}{\rho_0 L^2} - \frac{g}{T_0}\frac{dT_0}{dx_0} - - \frac{g}{2p_0}\frac{dB^2_0}{dx_0}.
\label{Gamma2}\end{eqnarray}
This form demonstrates that instability is driven in a magnetised atmosphere by temperature decreasing upwards and magnetic field decreasing upwards (magnetic buoyancy).   Note that in our case with thermal conduction along the field it is the temperature, not entropy gradient, that matters -- see discussion in {\em e.g.} Balbus \cite{balbus}).  The coefficients $C_0,\;C_2,\;C_3,\;C_4$ are given by:
\begin{eqnarray}
C_0 = \left(1 +\frac{\rho_0^2g^2L^2}{p_0^2\pi^2}\right)  &\;\;\;\;&
C_2 = -\left(\frac{B_0^2 \pi^2}{\rho_0 L^2}\right) \nonumber \\
C_3 = \left(\frac{B_0^2 \pi^2}{8\rho_0 L^2}\right) &\;\;\;\;&
C_4 = \frac{4}{3\pi}\left(\frac{g}{\rho_0}\frac{d^2\rho_0}{dx_0^2} - \frac{\rho_0^2g^3}{p_0^2}\right)
\label{coeff}\end{eqnarray}
Note that these coefficients have a different normalisation to the corresponding coefficients in \cite{cowart} -- but we have kept the same notation.  From Eq.~(\ref{nonlineareq}) we can derive an energy equation:
\begin{eqnarray}
\frac{d {\cal E}}{dt}  &=& \frac{d}{dt} \int dV\frac{1}{2}\left(C_0 \left(\frac{\partial\xi}{\partial t}\right)^2 - \Gamma^2(x_0)\xi^2 - C_2\left(\frac{\partial^2 u}{\partial x_0\partial y_0}\right)^2 + C_3\frac{1}{2}\left(\frac{\partial{\overline {\xi^2}}}{\partial x_0}\right)^2 - C_4 \frac{2}{3}\xi^3\right) \nonumber \\
 &=& -\nu\int dV\left(\frac{\partial^2\xi}{\partial t\partial y_0}\right)^2.
\label{energyC}\end{eqnarray}
where the integrals are taken over the whole plasma volume and $dV= dx_0dy_0L$.  The total energy ${\cal E}$ is obviously of the form of kinetic energy and a nonlinear potential energy.  Note that motion can be only be driven by the terms that lower the potential energy.  In this case these are the linear drive in regions where $\Gamma_0^2(x_0)>0$ and the nonlinear drive where $C_4 \frac{2}{3}\xi^3 >0$.
The inertial term ($C_0$) arises from both the vertical acceleration and the acceleration along the tilted field lines.  The stabilising $C_2$ linear term arises from the sideways, y, displacement  bending the field lines.  The $C_3$ stabilising quasilinear term arises from the modifications of the mean profile by the perturbations.    The $C_4$ explosive nonlinearity is examined in the flux tube limit below.  We presume that the system evolves through marginal stability -- perhaps by a slow lengthening of the box $L$.   Then in some narrow region around the height $x_{max}$ the system is just above local marginal stability $\Gamma_0^2(x_{max}) \sim {\cal O}(n^{-1})\frac{B_0^2 \pi^2}{\rho_0 L^2}$ and that everywhere else it is locally stable $\Gamma_0 <0$.  In the unstable region we can approximate:
\begin{equation}
\Gamma_0^2 = \Gamma_{0}^2(x_{max}) - \left|\frac{d^2\Gamma_0^2}{dx^2}\right| \left(\frac{(x_0 -x_{max})^2}{2}\right).
\end{equation}
Thus the unstable region has the width $x_0 - x_{max} = 2\left(\frac{1}{2\Gamma_0^2}|\frac{d^2\Gamma_0^2}{dx^2}|\right)^{-1/2}$.  Since we expect the perturbation to be localised around the unstable region we may replace the smoothly varying functions $C_0\; ......C_4$ by constants given by the functions evaluated at $x_{max}$.  The dynamics close to marginal stability yields the generic form Eq.~(\ref{nonlineareq}) for fine scale MHD instabilities (with complicated expressions for the constants $C_0\; ......C_4$) see \cite{HFC, wilcow}.  The dynamical behaviour of Eq.~(\ref{nonlineareq}) was examined in detail in \cite{cowart, fong, fong2, cowartbright} -- we provide only a brief summary.

\subsection{Linear Eigenfunctions.}
\label{linear}

For very small amplitude, $\xi \ll \frac{\Gamma_{0}^2}{C_4}$, the motion is linear and the most unstable eigenfunctions for a given $k_y$ are given by:
\begin{equation}
\xi = \xi_0 e^{[\gamma t - \frac{(x-x_{max})^2}{2\Delta^2} +ik_y y_0]}
\end{equation}
where the width and growth rate are given by: 
\begin{eqnarray}
\Delta & = & \left(  \frac{2|C_2|}{k_y^2|\frac{d^2\Gamma_0^2}{dx^2}|}\right)^{1/4} \\ \nonumber \\
\gamma & = & \sqrt{\frac{1}{C_0}\left(\Gamma_0^2 - \frac{\sqrt{\frac{1}{2}|C_2||\frac{d^2\Gamma_0^2}{dx^2}|}}{|k_y|}+ \frac{\nu^2k_y^4}{4C_0}\right)} -\frac{\nu k_y^2}{2C_0}.
\end{eqnarray}
The threshold for instability at any $k_y$ is $\Gamma_0^2>0$.  When this is true the growth rate is positive (instability) for $k_y^2 >k_c^2 = \frac{1}{2\Gamma_0^4}|C_2||\frac{d^2\Gamma_0^2}{dx^2}|$.  For weak viscosity, $\nu k_c^2 \ll \Gamma_0$, the peak growth rate is $\gamma_{max} \sim \Gamma_0/\sqrt{C_0}$ and the wave number at the peak is $k_{peak} \sim \frac{1}{\nu^{1/3}}\left(\frac{1}{8\Gamma_0^2}C_0|C_2||\frac{d^2\Gamma_0^2}{dx^2}|\right)^{1/6}$.  The eigenfunction and growth rate are consistent with the assumed scalings -- {\em i.e.} $\gamma \sim  {\cal O}(\Gamma_0) \sim {\cal O}(n^{-1/2})\Gamma_A$, $k_y \sim {\cal O}(n)\frac{1}{L_\gamma}$ and $\Delta \sim {\cal O}(n^{-1/2}) L_\gamma$ where the Alfv\'{e}n frequency is $\Gamma_A = \frac{B_0 \pi}{\sqrt{\rho_0} L}=\sqrt{|C_2|}$ and the scale length is defined by $L_\gamma^2 = \Gamma_A^2/|\frac{d^2\Gamma_0^2}{dx^2}|$.  The linear modes, as expected, are localised in $x_0$ and fine scale in $y_0$.

\subsection{Nonlinear Behaviour.}
\label{nlinear}

\begin{figure}[h]
\setlength{\unitlength}{1cm}
\begin{center}
{\includegraphics[angle=0, width=11.0cm, totalheight=9.0cm,trim=0 0 0 0,clip]{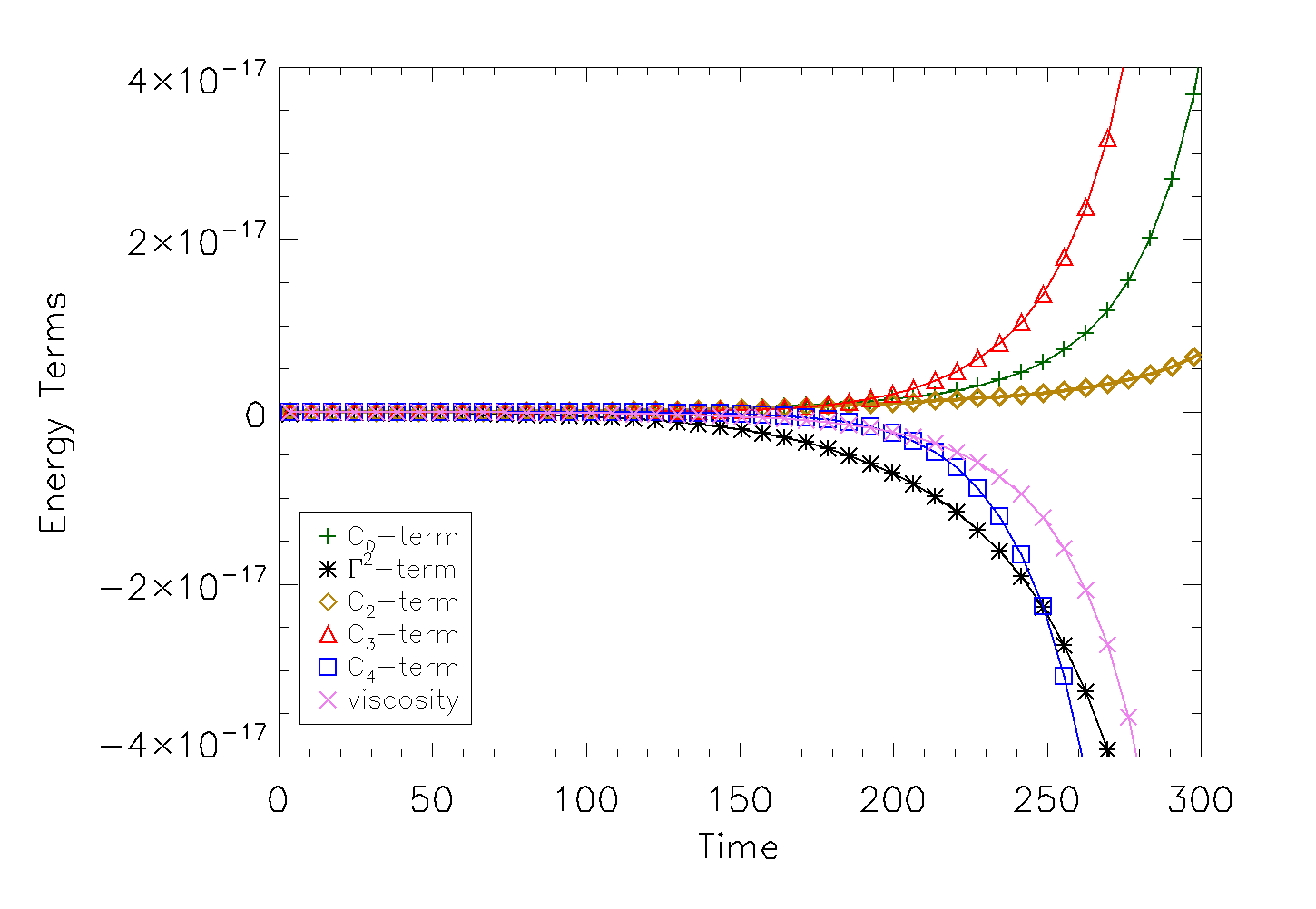}
}
\vskip -0.3 truein
\caption{\textit{Contributions to energy versus time in early evolution of most unstable perturbation  of Eq.~(\ref{nonlineareq}). Lines correspond to terms in Eq.~(\ref{energyC}).  Note that until about $t=210$ the linear drive $\Gamma^2$ dominates the release of negative energy to drive the instability. However by $t=250$ the explosive nonlinearity (the $C_4$ term) dominates the drive.  Note the energy is very small when nonlinearity becomes important because the linear drive is very small.}} 
\label{energy1}\end{center}\end{figure}

\begin{figure}[h]
\setlength{\unitlength}{1cm}
\begin{center}
{\includegraphics[angle=0, width=11.0cm, totalheight=9.0cm,trim=0 0 0 0,clip]{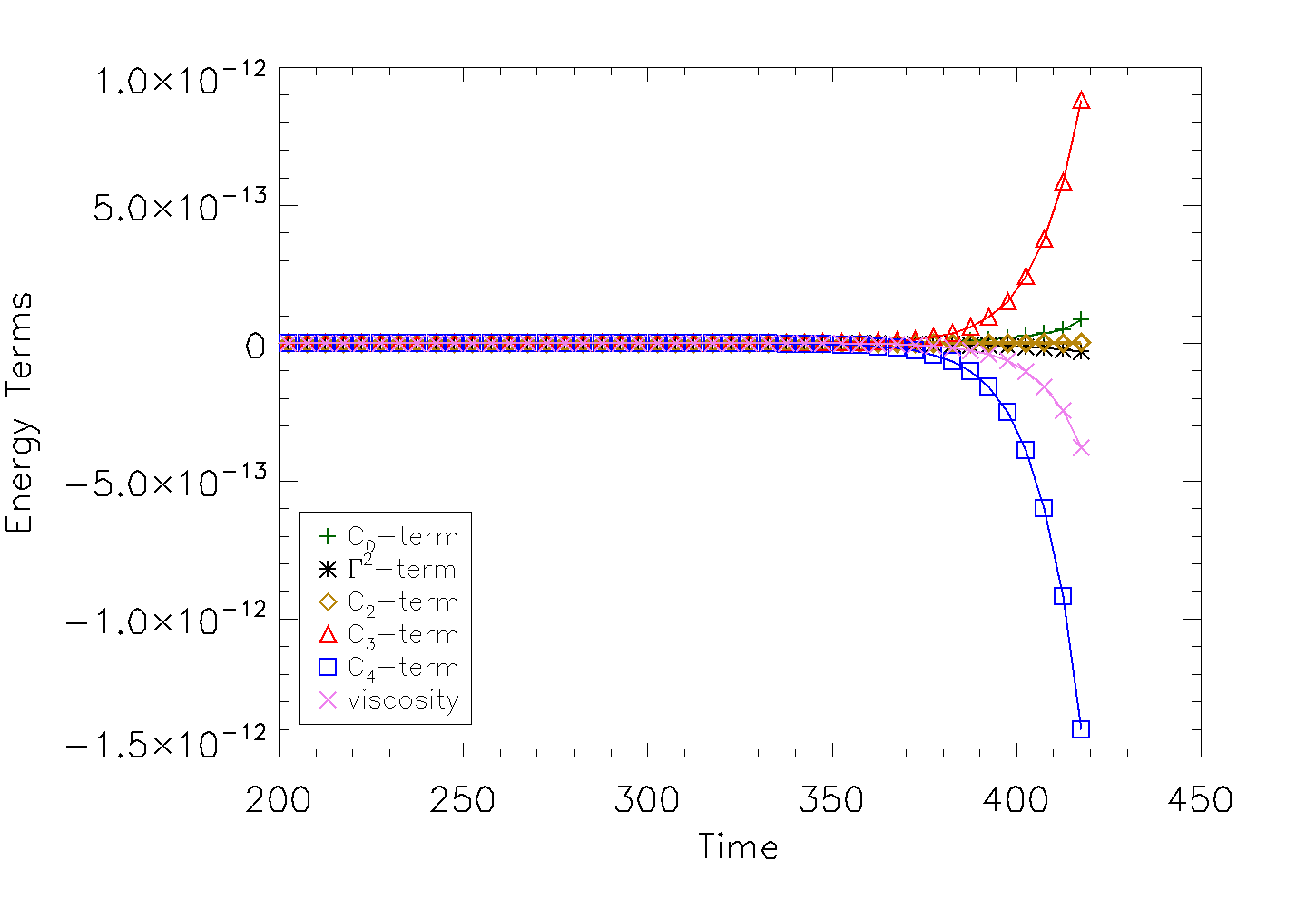}
}
\vskip -0.3 truein
\caption{\textit{Contributions to energy versus time in late evolution of most unstable perturbation  of Eq.~(\ref{nonlineareq}). Lines correspond to terms in Eq.~(\ref{energyC}).  The solution is approaching a singularity at $t = t_0 \sim 470.5$.  The singular solution is clearly dominated by a balance between the two nonlinear terms (the $C_3$ and $C_4$ terms). }} 
\label{energy2}\end{center}\end{figure}

\begin{figure}[h]
\setlength{\unitlength}{1cm}
\begin{center}
{\includegraphics[angle=0, width=9.0cm, totalheight=8.0cm,trim=1 0 0 0,clip]{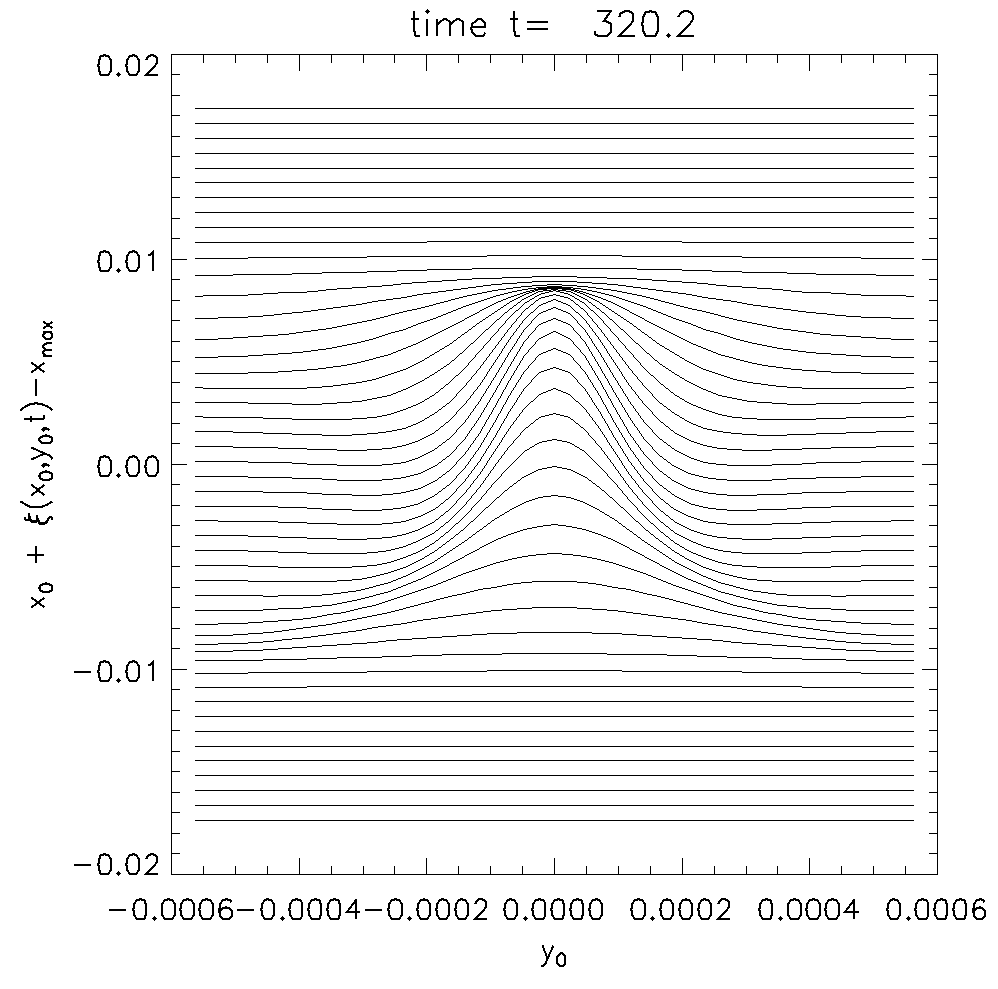}
}
\vskip -0.2 truein
\caption{\textit{Lines of displaced plasma height $x - x_{max}$ (i.e. the displaced flux surfaces) for fixed initial height $x_0$ where $x = x_0 + \xi(x_0, y_0, t)$.  $\xi(x_0, y_0, t)$ is obtained from a numerical solution to Eq.~(\ref{nonlineareq}) with coefficients given by Eq.~(\ref{modelcoef}) at time $t=320.2$.  This solution is deep into the nonlinear phase of evolution and the surfaces are close to overtaking.  Note the difference in scales of $x_0$ and $y_0$ -- the finger is much narrower than it appears.}} 
\label{finger}\end{center}\end{figure}

The nonlinear terms in Eq.~(\ref{nonlineareq}) become important  when $\xi \sim \frac{\Gamma_{0}^2}{C_4}$ for the explosive nonlinearity and $\xi^2 \sim \frac{\Gamma_{0}^2\Delta^2}{C_3}$ for the quasilinear nonlinearity.  For $C_4>0$ the explosive nonlinearity  drives the motion in the upwards (x increasing) direction and stabilises the downwards motion.  We shall assume $C_4>0$ for discussion here since the $C_4<0$ case can be obtained by reversing the sign of $\xi$.  The $C_3$ nonlinearity tries to reduce the $y$ average motion ({\em i.e.} ${\overline {\xi^2}}$) and broaden the mode in $x$; {\em i.e.} minimise the flattening of the mean profiles.  Numerical solution of Eq.~(\ref{nonlineareq}) (\cite{cowart, fong, fong2, cowartbright, wilcow}) reveals a generic scenario for the small amplitude nonlinear evolution.  First the instability grows linearly in the unstable region.   Once the amplitude is large enough that the nonlinear terms (the $C_3$ and $C_4$ terms) begin to dominate, the upward moving plasma accelerates and narrows in y.  The dynamics becomes characterised by fingers of explosively growing upwards moving plasma of width $\Delta y$ -- see Figure~(\ref{finger}).  The evolution approaches a finite time singularity -- see \cite{cowart, fong, fong2, cowartbright, wilcow}.  Here we present an example numerical solution of Eq.~(\ref{nonlineareq}) with coefficients derived from the model of Section (\ref{numerical}).  This solution is calculated in the normalised tilde variables of Section (\ref{numerical}) where the coefficients (which are the same for both CASE 1. and 2.) are
\begin{eqnarray}
{\Gamma}^2(\tilde{x}_0) \sim 1.9\times 10^{-4} - 1.31\frac{({x}_0 - {x}_{max})^2}{2} \nonumber \\
{C}_0 = 0.248 \;\;\;\;  {C}_2 = - 0.352 \;\;\;\;  {C}_3 = 0.044 \;\;\;\;  {C}_4 = 0.216 \;\;\;\; 
\label{modelcoef}\end{eqnarray}
and ${\tilde x}_{max} \sim 1.1118$.  Note that the system is weakly growing over a very narrow region ($({\tilde x}_0 - {\tilde x}_{max}) \sim 0.017$).  We take $\nu = 10^{-10}$ and initialise with the most unstable linear mode with $k_y = 5412$ -- {\em i.e.} a wavelength of $\lambda_y = 0.00116$.  In Figures~(\ref{energy1}) and  (\ref{energy2}) we plot the time behaviour of the energy terms of Eq.~(\ref{energyC}) at early and late times.  The simulation is terminated at a time of $t=418$ when the resolution becomes too poor.
\begin{figure}[h]
\setlength{\unitlength}{1cm}
\begin{center}
{\includegraphics[angle=0, width=7.5cm, totalheight=5.5cm,trim=0 0 0 0,clip]{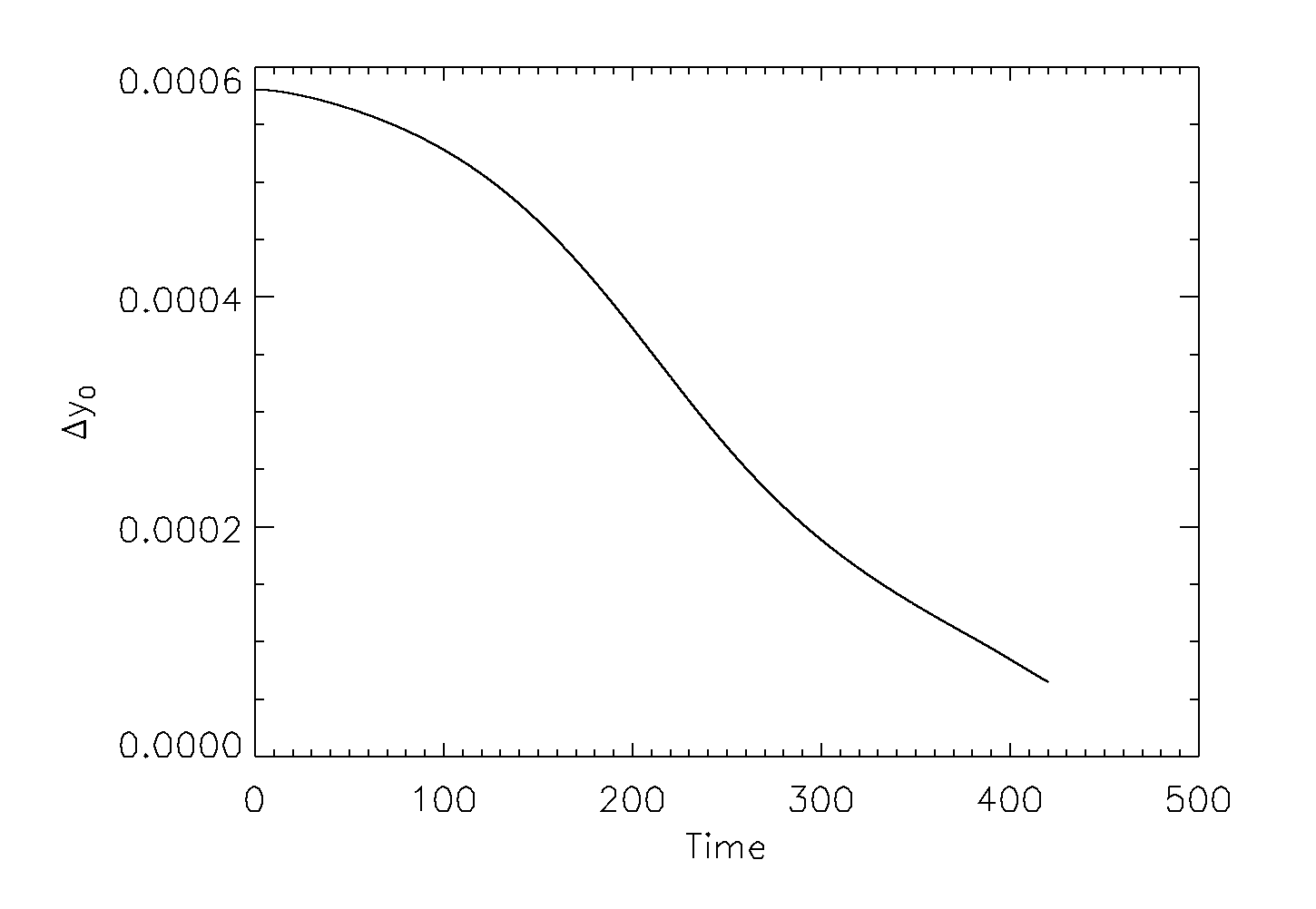}
}{\includegraphics[angle=0, width=7.5cm, totalheight=5.5cm,trim=0 0 0 0,clip]{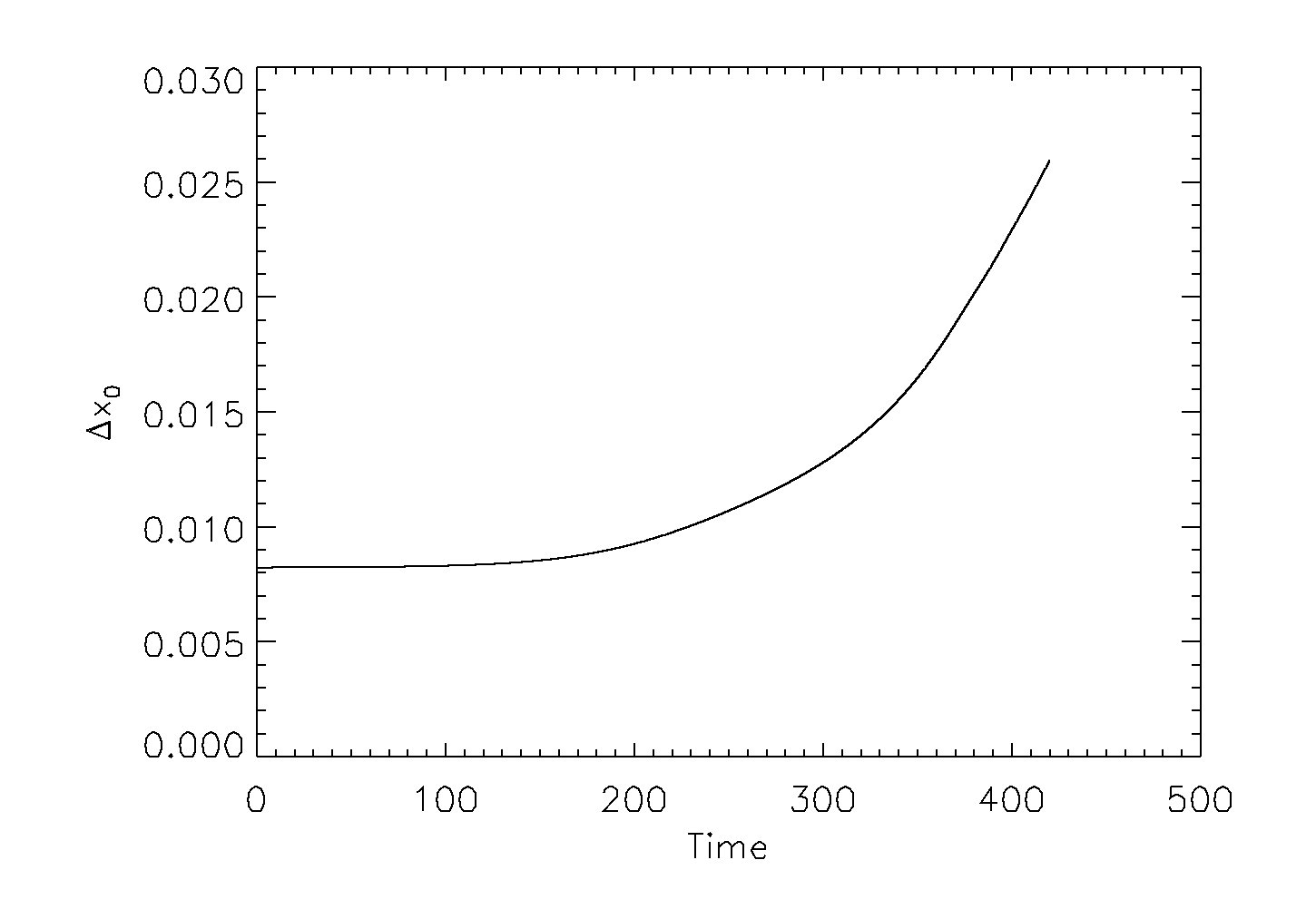}
}
\vskip -0.2 truein
\caption{\textit{Time evolution of the finger width $\Delta y_0$ (left) and  $\Delta x_0$ (right), see definition in Eq.~(\ref{Deltay}).  These measures show the narrowing of the unstable finger in $y_0$ and the spreading (in $x_0$) of the perturbation into the metastable region.}} 
\label{deltay}\end{center}\end{figure}

We define the "finger" width in y, $\Delta y_0$, and the disturbed height, $\Delta x_0$ by:  
\begin{eqnarray}
\Delta y_0 = \frac{\int_{-\lambda_y}^{\lambda_y}dy_0[\xi(x_{max}, y_0, t)]^2}{[\xi(x_{max}, 0, t)]^2} \;\;\;\;\; (\Delta x_0)^2 = \frac{\int dx_0dy_0(x_0 - x_{max})^2\xi^2}{\int dx_0dy_0\xi^2}.
\label{Deltay}\end{eqnarray}
For the evolution in Figures~(\ref{energy2}) and (\ref{deltay}) we can fit the later stage nonlinear motion to:
\begin{eqnarray}
\xi \sim (t - t_0)^{-2.52} \;\;\;\;\;\; \Delta y_0 \sim (t - t_0)^{1.04} \;\;\;\;\; \Delta x_0\sim (t - t_0)^{-0.73},
\label{scalingex}\end{eqnarray}
with $t_0 = 470.5$.  In this asymptotic regime the contribution of the linear terms ($C_1$ and $C_2$ terms) is negligible and the explosive nonlinearity (the $C_4$ term) drives the growth of kinetic energy against viscous dissipation and the stabilising quasilinear term (the $C_3$ term).  In the simulation Figure~(\ref{energy2}), above it is clear that the nonlinear terms largely balance with the remaining drive providing the viscous heating and growth of kinetic energy.  The dynamics depends on the 
form of the physics that limits the $k_y$. We have not been able to find a simple analytic derivation of the scaling in Eq.~(\ref{scalingex}) -- {\em i.e.} with viscosity.  However the balance of the two nonlinear terms ($C_3$ and $C_4$) yields $\frac{(\Delta x_0)^2}{\Delta y_0} \sim \xi$ which is obeyed by the scaling in Eq.~(\ref{scalingex}).  In \cite{fong, fong2} the transition to explosive growth with finite larmor radius terms providing the $k_y$ limit was examined.  Analytic expressions for the exponents of explosive growth were derived in this case.  These are, of course, different to those with viscous growth given in Eq.~(\ref{scalingex}).

We note that the (quasilinear) nonlinearity broadens the mode into the linearly stable region $x-x_{max}>\Delta$ -- field lines in this region are metastable so that when knocked hard by the rising finger of plasma they are destabilised.  This mechanism of progressive destabilisation was termed {\em detonation} \cite{cowart, fong, fong2, cowartbright, wilcow} because of the analogy with chemical explosives.  Linear instability is not necessary for explosive growth since finite perturbations in Eq.~(\ref{nonlineareq}) will destabilise linearly stable profiles and trigger detonation.  The quasilinear nonlinearity suppresses all but the largest amplitude fingers \cite{fong2}.  Thus by the end of this small amplitude evolution the dynamics consists of a few rapidly rising fingers of plasma.

The treatment in this section cannot capture displacements as large as the width of the unstable region since $\xi \sim {\cal O}(n^{-1})L_\gamma\ll {\cal O}(n^{-1/2})L_\gamma\sim \Delta$. Thus while the asymptotic regime is reached before the equations break down, the singularity itself is not.  What then happens to the rising fingers of plasma?  In the next section we examine a scenario for the further evolution of exploding fingers of plasma.

\section{Flux Tube Dynamics}
\label{fluxtube}
\begin{figure}[ht]
\setlength{\unitlength}{1cm}
\begin{center}
{\includegraphics[angle=0, width=16.0cm, totalheight=14.0cm, trim=0 0 0 0, clip]{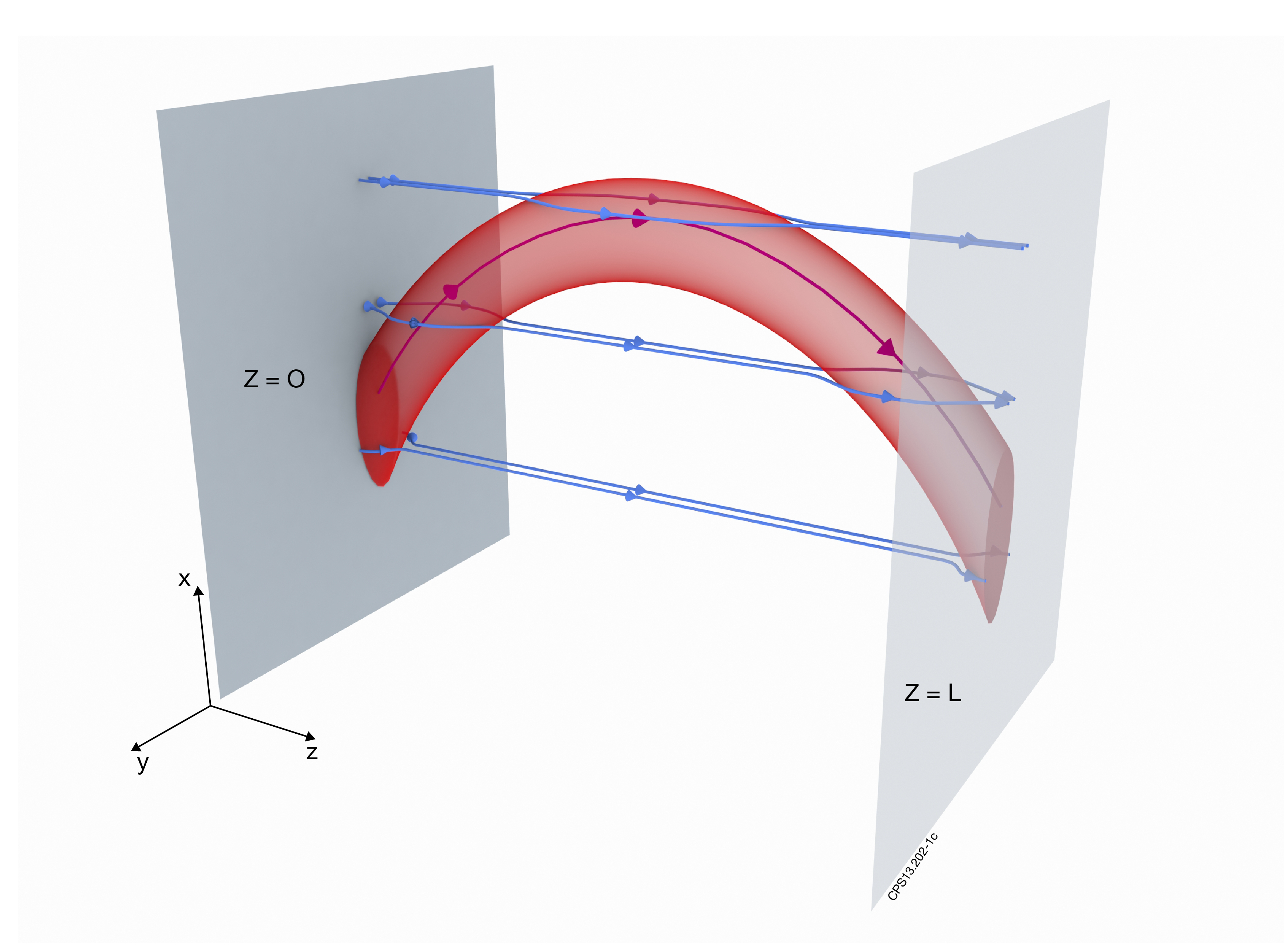}}
\vskip -0.3 truein\caption{\textit{Erupting flux tube pushing through the field above.  The tube is narrow to minimise the sideways (y) bending the external field lines.  The field lines are tied to the walls at $z=0, L$.
}} \label{erupt1}\end{center}\end{figure}

In this section we examine the finite amplitude dynamics of single isolated narrow line tied flux tubes in our box equilibrium.   The tubes have elliptical cross sections, elongated in the direction of motion ($x$) and narrower across ($\Delta y = \delta_1 \ll \Delta x = \delta_2\ll L$), see Fig.~(\ref{erupt1}).  The exact cross sectional shape of the tube is not important here -- just that it is narrow and considerably elongated in the direction of motion (see Fig.~(\ref{erupt1})).  This shape allows the tube to "knife" through the plasma separating the surrounding field lines very little -- indeed we shall take $\delta_1$ to be sufficiently small that to lowest order the surrounding field lines are effectively unperturbed.  We conjecture that such tubes are the later stage evolution of the fingers seen in the early stage nonlinear development described above.  The rising fingers in the small amplitude motion are, however, never independent (isolated).  Thus we must assume that as they evolve from fingers to a moving flux tube they become independent.  The isolated tubes are presumed to move somewhat slower than the sound speed since we are interested in the partially viscous behaviour near marginal stability and the saturated states of the flux tube.  

\begin{figure}[h]
\setlength{\unitlength}{1cm}
\begin{center}
{\includegraphics[angle=0, width=8.0cm, totalheight=9.0cm,trim=0 0 0 0,clip]{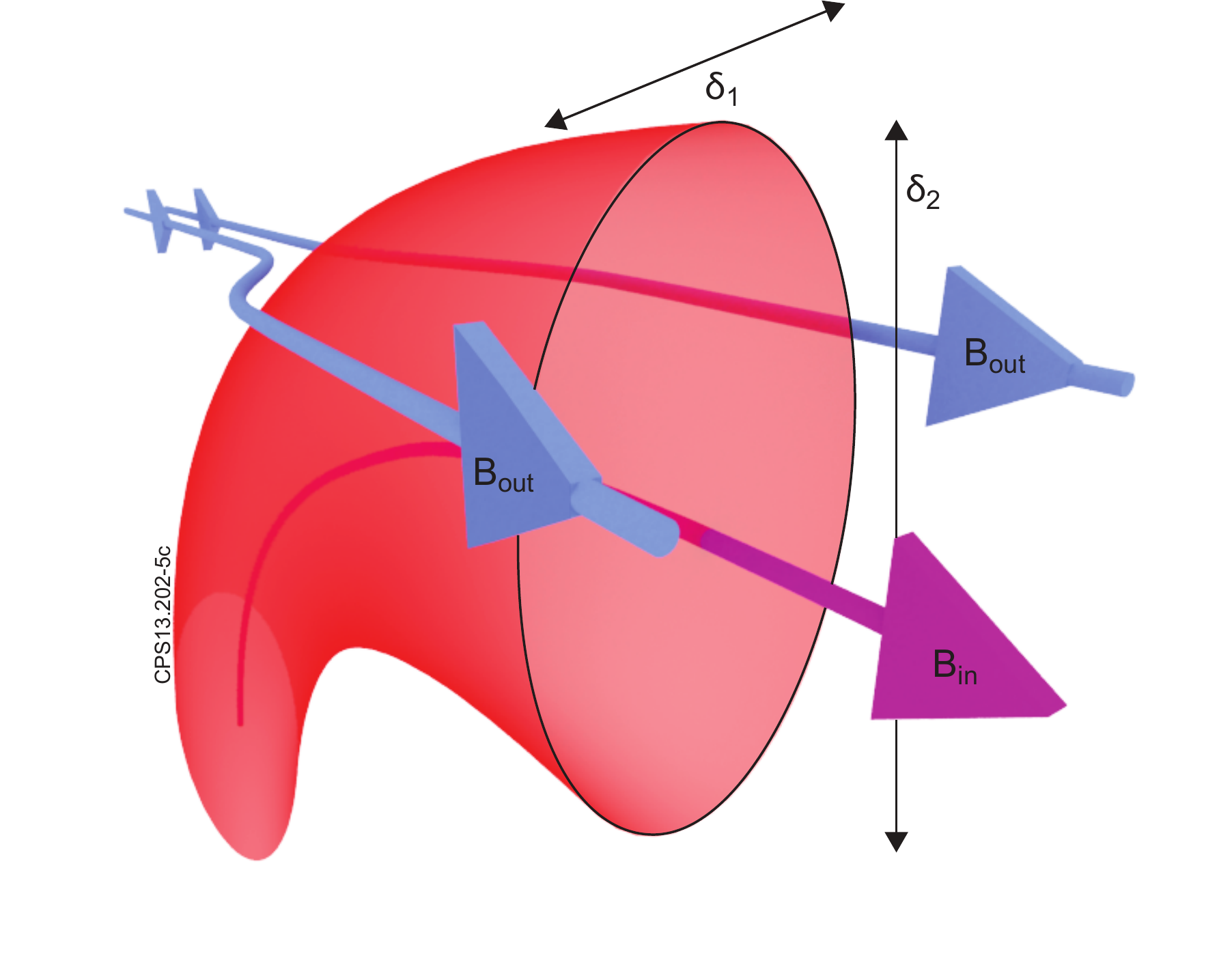}
}
\caption{\textit{The bent elliptical filament (red with mauve field line) pushing aside the (blue) surrounding field lines. The tube is assumed highly elliptical with $\delta_1\ll\delta_2$. Total pressure inside filament, $p_{in} + {B^2_{in}}/{2}$, is equal to the total pressure just outside the filament, $p_{out} + {B^2_{out}}/{2}$, at every point along the filament.
}} \end{center}\label{tube}\end{figure}

Consider the field aligned flux tube that is displaced through the plasma, see Fig.~(\ref{erupt1}).  The field inside the tube is denoted ${\bf B}_{in}$ and the field outside ${\bf B}_{out}$ -- see Fig.~(\ref{tube}).  The field in the tube is hardly bent in the $y$ direction, $B_y \sim {\cal O}( \delta_1/L)B_z$.  Thus the equation for a field line in the tube is given (to leading order) by $x=x(x_0, y_0, z, t)$ and $y=y_0$ where $x_0$ is the undisplaced height of the field line and $y_0$ is the undisplaced y position of the field line.  Since the $y$ dependence is not used further in this section we omit the $y_0$ dependence of $x$ in subsequent expressions, {\em i.e.} we write $x=x(x_0, z, t)$. The field lines are tied to the wall therefore $x(x_0, 0, t) = x(x_0, L, t) = x_0$.  We can write:
\begin{equation}
{\bf B}_{in} = B_z[ {\bf\hat z} +  \left(\frac{\partial x}{\partial z}\right)_{x_0}{\bf\hat x}] 
\label{Bin1}\end{equation}
where $B_z$ is a function of $x$ and $x_0$ to be found and $\bf\hat z$ and $\bf\hat x$ are unit vectors in the z and x direction respectively.  

The force (per unit volume) on the plasma is: 
\begin{equation}
{\bf F} = -\grad [p + \frac{B^2}{2}] + {\bf B}\cdot\grad {\bf B} - \rho g {\bf\hat x}.
\label{force}\end{equation}
We shall refer to $p + \frac{B^2}{2}$ as the {\em total pressure} and ${\bf B}\cdot\grad {\bf B}$ as the {\em curvature force}.  
The forces across the narrow tube (in the ${\bf\hat y}$ direction) are formally large, ${\cal O} (p/\delta_1)$, and must balance to this order {\em i.e.}
\begin{equation}
{\bf F}\cdot{\bf\hat y} \sim - \frac{\partial}{\partial y} [p + \frac{B^2}{2}] =0.
\label{forcey}\end{equation}
Thus fast waves propagate (at speed $C_{fast} = \sqrt{(p_0 + B_0^2)/\rho_0}$) across the tube on a time $\delta_1/C_{fast}$ and equalise the total pressure inside and outside the tube.  Thus on the slow evolution time:
\begin{equation}
p_{in} + \frac{B_{in}^2}{2} = p_{out} + \frac{B_{out}^2}{2},
\label{total}
\end{equation}
where "$in$" refers to inside the tube and "$out$" refers to just outside the tube (at the same $x$ and $z$ along the tube -- see Fig.~(\ref{erupt1})).  We will assume that the field and pressure outside the tube are unperturbed so that: 
\begin{equation}
 p_{out}(x) = p_0(x) \;\;\;\; and \;\;\;  B_{out}(x) = B_0(x)
\label{pressure_out}\end{equation}
are known. The total pressure forces at a point on the tube are thus identical to the total pressure forces on the plasma it replaced.  The field line bending of the external field lines (blue lines in Fig.~(\ref{tube})) gives a sideways $y$ force of order $B_{out}^2(\delta_1/\delta_2^2)$ where we have estimated that for a finitely displaced flux tube the external field line is bent a distance $\delta_1$ over a length of $\delta_2$.  Inserting this estimate in  Eq.~(\ref{forcey}) would give corrections to the internal total pressure of order $\delta (p_{in} + \frac{B_{in}^2}{2}) =B_{out}^2(\delta_1^2/\delta_2^2)$.  We will ignore the contribution of these corrections to the vertical $x$ forces (see Eq.~(\ref{curve})) see below.  

The temperature is constant along the flux tube and at $z= 0, L$ is equal to $T_0(x_0)$.  Thus $T(x,z,t) = T_0(x_0)$ where $x_0 = x_0(x,z,t)$ is obtained from inverting $x=x(x_0, z, t)$.  Since we are interested in stable displaced equilibrium states of the flux tube and slow eruptions from an unstable state we set ${\bf F\cdot B} = 0$.
Thus:
\begin{eqnarray}
{\bf B}_{in}\cdot\grad p_{in} =  \frac{T_0(x_0)}{m}{\bf B}_{in}\cdot\grad \rho_{in} = -\rho_{in}g({\bf B}_{in}\cdot {\bf\hat x} ) \nonumber \\
\rightarrow \;\; \rho_{in}(x, x_0) = \rho_0(x_0)e^{-\left(\frac{mg(x-x_0)}{T_0(x_0)}\right)}
\label{pressure_in}\end{eqnarray}
where we have used the boundary condition that $x=x_0$ at $z= 0, L$.  The pressure is given by $p_{in}(x, x_0) = \rho_{in}(x, x_0) \frac{T_0(x_0)}{m}$.

From Eqs.~(\ref{pressure_in}), (\ref{pressure_out}), (\ref{total}) and using the equilibrium relation Eq.~(\ref{equilib})we obtain:
\begin{eqnarray}
B^2_{in}(x, x_0) = B_0^2(x_0) + 2p_0(x_0)\left[1 - e^{-\left(\frac{mg(x-x_0)}{T_0(x_0)}\right)}\right] - 2\int_{x_0}^{x}g\rho_0(x')dx'
\label{B_intot}\end{eqnarray}
Note that:
\begin{eqnarray}
\frac{1}{2}\left(\frac{\partial B^2_{in}}{\partial x}\right)_{x_0} = g[\rho_{in}(x, x_0)- \rho_0(x)]
\label{ident}\end{eqnarray}

Therefore the force in the tube in the direction of motion $x$:
\begin{eqnarray}
{\bf F}\cdot {\xx} = F_x & = & [{\bf B}_{in}\cdot\grad {\bf B}_{in} - \grad (p_{in} + \frac{B_{in}^2}{2}) - g\rho_{in}(x, x_0)\xx]\cdot {\xx} \nonumber \\ \nonumber \\ & = & [{\bf B}_{in}\cdot\grad {\bf B}_{in} - \grad (p_{out} + \frac{B_{out}^2}{2}) - g\rho_{in}(x, x_0)\xx]\cdot {\xx}\nonumber \\ \nonumber \\ & = & [{\bf B}_{in}\cdot\grad {\bf B}_{in} + (g\rho_0(x) - g\rho_{in}(x, x_0))\xx]\cdot {\xx}
\label{curve}
\end{eqnarray}
where we have used Eqs~(\ref{total}) and (\ref{pressure_out}) and equilibrium force balance Eqs~(\ref{equilib}).  Note that the force is simply {\em the field line bending force and the buoyancy force from Archimedes' principle.}   We have ignored the vertical forces from the corrections to the total pressure ($\delta (p_{in} + \frac{B_{in}^2}{2}) =B_{out}^2(\delta_1^2/\delta_2^2)$) in  the second line of Eq.~(\ref{curve}).  This requires 

\begin{equation}
\frac{\partial}{\partial x}\delta (p_{in} + \frac{B_{in}^2}{2}) \sim \frac{\delta (p_{in} + \frac{B_{in}^2}{2})}{\delta_2} \sim B_{out}^2\frac{\delta_1^2}{\delta_2^3}\ll {\bf B}_{in}\cdot\grad {\bf B}_{in}\cdot {\xx} \sim \frac{B_{in}^2 (x-x_0)}{L^2}.
\label{elliptical}\end{equation}

Thus (with $B_{in} \sim {\cal O}(B_{out})$) the ellipticity of the flux tube must exceed a critical value for the theory of this section to hold: 

\begin{equation}
\frac{\delta_1}{\delta_2}\ll  \sqrt{\frac{\delta_2(x-x_0)}{L^2}}.
\label{elliptical2}\end{equation}

This relation is obeyed for a finitely displaced flux tube ($(x-x_0)\sim L$) with the cross sectional shape of the weakly nonlinear regime (the left hand side of Eq.~(\ref{elliptical2}) is ${\cal O}(n^{-1/2})$ and the right hand side  is ${\cal O}(n^{-1/4})$).  During detonation $\delta_2$ increases and $\delta_1$ decreases so the inequality in Eq.~(\ref{elliptical2}) becomes even better satisfied.  Thus by the time the flux tube is finitely displaced we expect the inequality to be well satisfied.  \footnote{In the weakly nonlinear regime the dominant vertical forces (those in Eq.~(\ref{curve}))  cancel to lowest order due to the marginal stability.  Therefore the corrections to the force due to the sideways (y) motion {\em are} kept in Eq.~(\ref{nonlineareq}).}

Since Eq.~(\ref{curve}) only involves derivatives along the field line we can consider dynamics of each field line separately.  Using Eqs.~(\ref{B_intot}), (\ref{ident}) and (\ref{Bin1}) in Eq.~(\ref{curve}) we obtain:
\begin{eqnarray}
F_x & = & \frac{B^2_{in}}{[1+ (\frac{\partial x}{\partial z})_{x_0}^2]^2}\left( \frac{\partial ^2x}{\partial z^2}\right)_{x_0} - \frac{\frac{1}{2}(\frac{\partial B^2_{in}}{\partial x})_{x_0}}{[1+ (\frac{\partial x}{\partial z})_{x_0}^2]}
\nonumber \\ \;\;\; \nonumber \\
& = & -\frac{1}{2 (\frac{\partial x}{\partial z})_{x_0}}\left(\frac{\partial}{\partial z}\left(\frac{B^2_{in}}{[1+ (\frac{\partial x}{\partial z})_{x_0}^2]}\right)\right)_{x_0}
\label{nonlinFx}
\end{eqnarray}

Equilibrium position of the field lines satisfies $F_x = 0$ -- a second order nonlinear ordinary differential equation for $x = x(x_0, z, t)$ for each $x_0$. We integrate $F_x = 0$ (using Eq.~(\ref{nonlinFx})) with the boundary conditions:
\begin{equation}
x(x_0, z=0) = x_0;\;\;\;\; B_{in}(x_0, x_0) = B_0(x_0)\;\;\;\; and \;\;\;\;  (\frac{\partial x}{\partial z})_{x_0} (z=0) = q
\end{equation}
to obtain:
\begin{equation}
\left(\frac{\partial x}{\partial z}\right)_{x_0} = \pm\sqrt{q^2 + (1+q^2)\frac{B^2 - B_0^2}{B_0^2}}.
\label{shoot}\end{equation}
The solution of this equation must satisfy the additional boundary condition that $x(x_0, z=L) = x_0$.  Therefore by symmetry the peak of the field line is then at $z=L/2$ with $x=x_{peak}$ and $(\frac{\partial x}{\partial z})_{x_0}=0$.  Thus $x_{peak}(q)$ must satisfy:
 \begin{equation}
\frac{q^2}{(1+q^2)} = \frac{B_0^2(x_0) - B_{in}^2(x_{peak}, x_0)}{B_0^2(x_0)}
\end{equation}
and $q$ must satisfy the eigenvalue condition:
\begin{equation}
\int_{x_0}^{x_{peak}(q)}\frac{dx}{\sqrt{q^2 + (1+q^2)\frac{B_{in}^2(x, x_0) - B_0^2(x_0)}{B_0^2(x_0)}}} = \frac{L}{2}.
\end{equation}
In practice it is simpler to solve Eq.~(\ref{shoot}) numerically by shooting and iterating $q$ until we find $(\frac{\partial x}{\partial z})_{x_0}=0$ at $z=L/2$.    In the next section we use this procedure to find the equilibria for a simple model atmosphere.  

The magnetic energy in a narrow flux tube with flux $d\psi = BdA$ is proportional to $\int B^2dV = d\psi \int Bdl$ where the $dl$ integration is a line integral along the flux tube.  We now show that the equilibria are stationary 
points of the magnetic energy.  Let us define the energy functional
\begin{equation}
{\cal E_B}(x(z, t), x_0) = \int B_{in}(x, x_0) dl = \int_0^L B_{in}(x, x_0)\sqrt{[1+ (\frac{\partial x}{\partial z})_{x_0}^2]} dz 
\label{energy}\end{equation}
where the $z$ integration is at fixed $x_0$.  Varying $x(z, x_0, t)$ in ${\cal E_B}$ keeping $x_0$ constant we get:
\begin{eqnarray}
\delta{\cal E_B} & = & \int_0^L \left( (\delta x)\frac{\partial B_{in}}{\partial x}\sqrt{[1+ (\frac{\partial x}{\partial z})_{x_0}^2]} + B_{in}\left(\frac{\partial \delta x}{\partial z}\right)_{x_0}\frac{(\frac{\partial x}{\partial z})_{x_0}}{\sqrt{[1+ (\frac{\partial x}{\partial z})_{x_0}^2]}} \right)dz \nonumber \\ \nonumber \\
& = & -\int_0^L (\delta x) F_x\frac{\sqrt{[1+ (\frac{\partial x}{\partial z})_{x_0}^2]}}{B_{in}} dz 
\label{stationary}\end{eqnarray}
where we have integrated by parts and used $(\frac{\partial B_{in}}{\partial z})_{x_0} = (\frac{\partial x}{\partial z})_{x_0}\frac{\partial B_{in}}{\partial x}$. Thus equilibrium, $F_x = 0$, is a stationary state of the energy functional.
To model the dynamics we take a simple drag to balance the force i.e.
\begin{eqnarray}
 \nu_D\left(\frac{\partial x}{\partial t}\right)_{x_0}  =  F_x 
 =  \frac{B^2_{in}}{[1+ (\frac{\partial x}{\partial z})_{x_0}^2]^2}\left( \frac{\partial ^2x}{\partial z^2}\right)_{x_0} - \frac{\frac{1}{2}(\frac{\partial B^2_{in}}{\partial x})_{x_0}}{[1+ (\frac{\partial x}{\partial z})_{x_0}^2]}.
\label{drag}\end{eqnarray}
This models much more complex viscous and aerodynamic drag on the moving flux tube.  The form of the drag does not, of course, affect the final erupted state and the explosive nature of the eruption.  Motion of the tube with drag reduces the energy (monotonically) since:
\begin{equation}
\frac{\partial{\cal E_B}}{\partial t} =  -\int_0^L \nu_D \left(\frac{\partial x}{\partial t}\right)^2_{x_0}\frac{\sqrt{[1+ (\frac{\partial x}{\partial z})_{x_0}^2]}}{B_{in}} dz < 0. 
\label{stationary}\end{equation}
Thus evolution takes the flux tube to an energy minimum.  These are linearly stable positions of the flux tube -- not necessarily a global minimum of the energy.  We emphasise that in our approximation the dynamics of each field line is independent.  We can therefore consider each $x_0$ separately. 

\subsection{Small Amplitude Flux Tube Motion.}
\label{small amp}

Here we examine the small amplitude motion of the flux tube to connect the flux tube theory to the more general small amplitude theory given above in Section~(\ref{pert}).  Let us assume $x(x_0, z, t) = x_0 + \Delta x$ with $\Delta x$ much smaller than the typical scale height.  Expanding Eq.~(\ref{drag}) to second order in $\Delta x$ we obtain:
\begin{eqnarray}
 \nu_D\left(\frac{\partial \Delta x}{\partial t}\right)_{x_0}  = B_0^2\left( \frac{\partial ^2\Delta x}{\partial z^2}\right)_{x_0} 
+\left(\rho_0\frac{mg^2}{T_0} + g\frac{d\rho_0}{dx_0}\right)(\Delta x)  \nonumber \\  + \frac{1}{2}\left(g\frac{d^2\rho_0}{dx_0^2} - g\rho_0(\frac{mg}{T_0})^2\right)(\Delta x)^2 + {\cal O}((\Delta x)^3)
\label{quasilin}\end{eqnarray}
where we have expanded the right hand side of Eq.~(\ref{ident}).  Close to marginal stability the first two terms on the right hand side of Eq.~(\ref{quasilin}) almost cancel for $\Delta x = \xi(t)\sin{(\frac{\pi z}{L})} + {\cal O}(\xi^2)$.  Substituting this into Eq.~(\ref{quasilin}) multiplying by $\sin{(\frac{\pi z}{L})}$ and integrating over $z$ from $0$ to $L$ we get:
\begin{eqnarray}
 \frac{\nu_D}{\rho_0}\left(\frac{d \xi}{dt}\right)_{x_0}  = \Gamma_0^2(x_0)\xi +  C_4\xi^2+ {\cal O}(\xi^3)
\label{ampevo}\end{eqnarray}
where $ \Gamma_0^2(x_0)$ is given in Eq.~(\ref{Gamma}) and $C_4$ in Eq.~(\ref{coeff}).  Clearly the small amplitude flux tube dynamics captures the linear and explosive nonlinearity parts of Eq.~(\ref{nonlineareq}) --  the assumption of thin isolated flux tubes orders out the remaining terms in the forces of (\ref{nonlineareq}) and drag replaces the inertial and viscous terms.  Again we will take $C_4 >0$ since the case $C_4 <0$ is reproduced by changing the sign of $\xi$. Let $\xi_0$ be the initial displacement, $\xi_c = \Gamma_0^2(x_0) /C_4$ and $\gamma = \frac{\Gamma_0^2(x_0)\rho_0}{\nu_D}$ the linear growth rate.  Then:
\begin{eqnarray}
\xi(t) = \frac{\xi_0e^{\gamma t}}{(1 + \frac{\xi_0}{\xi_c}) - \frac{\xi_0}{\xi_c}e^{\gamma t}}.
\label{ampevo1}\end{eqnarray}
In the linearly unstable region $\gamma >0$ displacements grow from infinitesimal amplitudes.  For $\xi_0>0$ the growth accelerates (explosively) to infinity in a finite time, $t_\infty = \frac{1}{\gamma}\ln {(1 + \frac{\xi_c}{\xi_0})}$; for  $\xi_0<0$ the growth saturates at an amplitude  $\xi = -\xi_c$.  In the linearly stable region $\gamma <0$ the displacement decays unless $\xi > -\xi_c$ when it grows explosively.  Thus, $-\xi_c$ (which is positive for $\gamma <0$) is the {\em critical amplitude} to excite a metastable field line.  Clearly the small amplitude dynamics is of the transcritical form with $A_c =
-\xi_c$.

\subsection{Flux Expulsion.}
\label{Expulsion}

In situations with substantial plasma beta ($\beta = p/B^2 \sim {\cal O}(1)$) the field lines can erupt until one point or part of the flux tube has zero magnetic field -- ({\em i.e.} $B_{in}^2 = B_0^2(x) + 2(p_0(x) - p_{in}(x))=0$).  The flux tube has then expanded to an infinite cross section to conserve the flux.  Clearly this breaks the assumptions of a thin isolated flux tube.  Indeed our theory must be restricted to cases with $B_{in} \gg (\delta_1/L)B_0$.  But we can still apply our theory when the field line approaches the $B_{in}^2 =0$ asymptotic limit closely if we consider $\delta_1 \rightarrow 0$.  Here we examine what happens in this limit.
\begin{figure}[h!]
\setlength{\unitlength}{1cm}
\begin{center}
{\includegraphics[angle=0, width=12.0cm, totalheight=7.0cm, trim=0 1cm 0 4cm, clip]{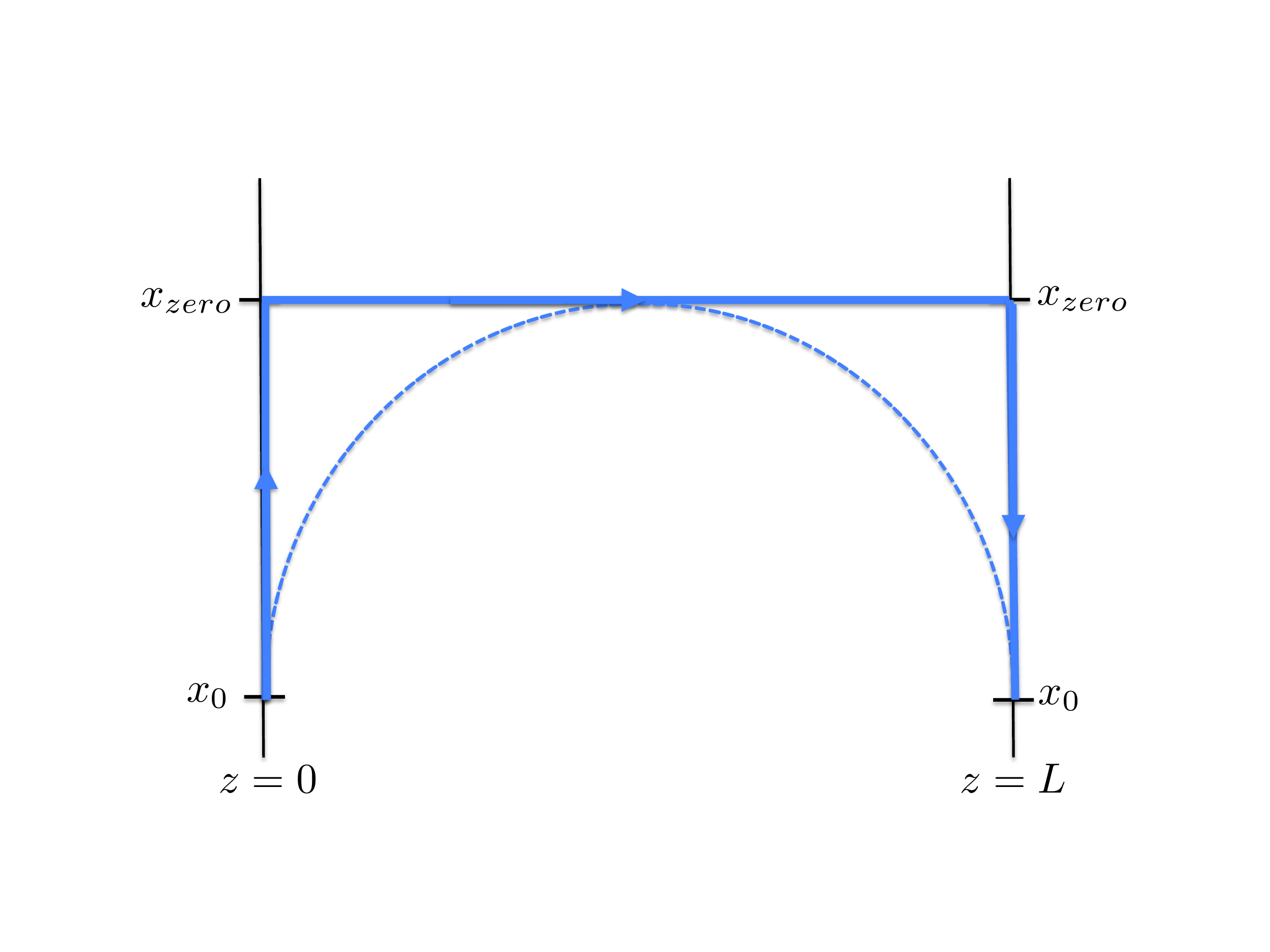}}
\caption{\textit{Field line initial shape illustrated by the dotted blue line where $x_{peak} = x_{zero}$ and $B_{in}(x_{zero}, x_0) =0$.  This line can lower its energy by evolving to the rectangular shape illustrated by the solid blue line -- {\em i.e.} made from the three segments: vertical from $z=0, \, x= x_0$ to $z=0, \, x= x_{zero}$, horizontal with $x = x_{peak} = x_{zero}$ for $0\leq z \leq L$ and vertical from $z=L, \, x= x_{zero}$ to $z=L, \, x= x_{0}$.}}.   \label{step}\end{center}\end{figure}

Suppose we have initially displaced the field line so that $x_{peak}= x_{zero}$ where $B_{in}(x_{zero}, x_0) =0$ -- see Fig.~(\ref{step}).  Then: 
\begin{equation}
{\cal E_B}(x(z, t), x_0) = \int B_{in}(x, x_0) dl =  2\int_{x_0}^{x_{peak}} B_{in}(x, x_0) (\frac{dl}{dx})dx. 
\label{energy0}\end{equation}
But $\frac{dl}{dx} \geq 1$ where the equality holds when the field line is vertical and $dl = dx$.  
Thus we can lower ${\cal E_B}(x(z, t), x_0)$ by making the field line have three segments: vertical from $z=0, \, x= x_0$ to $z=0, \, x= x_{zero}$, horizontal with $x = x_{peak} = x_{zero}$ for $0\leq z \leq L$ and vertical from $z=L, \, x= x_{zero}$ to $z=L, \, x= x_{0}$.  The horizontal segment makes no contribution to the energy since $B_{in}(x_{peak}, x_0) =0$.  We illustrate this minimum energy state with the continuous line in Fig.~(\ref{step}) -- note that the sharp corner in the field happens where $B_{in} =0$ and therefore has no {\em field line bending force}.  The minimum energy state has:
\begin{equation}
{\cal E_B}(x(z, t), x_0) = {\cal E_B}(x(z, t), x_0)_{min} =   2\int_{x_0}^{x_{peak}} B_{in}(x, x_0)dx. 
\label{energymin}\end{equation}
This is clearly a local minimum in energy -- not necessarily a global minimum.  However it is clear that if the field line peak reaches the point of zero field strength (the dotted line in Fig.~(\ref{step})) the motion will continue towards the minimum energy state.   The dynamics of the section of the field line with zero, or at least very small, field strength is a topic for future work.  It is clear, however, that the zero field section of the tube remains buoyant and will continue to rise.

\subsection{Flux Tube Motion in a Model Atmosphere.}
\label{model}

In this section we examine the flux tube motion in a simple model magnetized atmosphere.  Let the unperturbed density and magnetic field be given by:
 \begin{eqnarray}
\rho_0(x_0) = \frac{\bar{\rho_0}}{\cosh^2[(x_0 - x_{\rho})/L_\rho]} \nonumber \\
B_0^2(x_0) = \bar{B_1^2} - \frac{\bar{B^2_2}}{\cosh^2[(x_0 - x_{B})/L_\rho]} 
\label{definrhoB}\end{eqnarray}
where $\bar{\rho_0}$, $\bar{B_1^2}$ and $\bar{B^2_2}$ are constants and $x_{\rho}$ and $x_{B}$ are the heights of the maximum density and minimum field respectively.  From Eq.~(\ref{equilib}) we find:
 \begin{eqnarray}
p_0(x_0) = \bar{p_0} - \frac{1}{2}B^2_0(x_0) - g\bar{\rho_0}L_\rho \tanh [(x_0 - x_{\rho})/L_\rho]
\label{definp}\end{eqnarray}
where $\bar{p_0}$ is a constant.  We take large pressure to focus on the magnetised Rayleigh-Taylor instability (rather than the Parker instability) so that $\frac{mg L_\rho}{T_0(x_0)} = \frac{g\rho(x_0)L_\rho}{p_0(x_0)} \ll 1$.  We define normalised variables as:
 \begin{eqnarray}
\tilde{x} = \frac{x}{L_\rho},\;\;\;  \tilde{x_0} = \frac{x_0}{L_\rho},\;\;\;  \tilde{z} = \frac{z}{L},\;\;\; \tilde{t} = 2\frac{g\bar{\rho_0} t}{\nu_D L_\rho},\;\;\;  
\label{defintilde}\end{eqnarray}
and the constants
 \begin{eqnarray}
\tilde{x}_{B} = \frac{x_{B}}{L_\rho},\;\;\;  \tilde{x}_{\rho} = \frac{x_{\rho}}{L_\rho},\;\;\;\; A=\frac{L_\rho}{L},\;\;\;\;\;
{\tilde{B_1^2}} = \frac{\bar{B_1^2}}{2g\bar{\rho_0}L_\rho}, \;\;\;\; {\tilde{B_2^2}} = \frac{\bar{B_2^2}}{2g\bar{\rho_0}L_\rho}.
\label{consttilde}\end{eqnarray}
We also define a normalised flux tube field:
 \begin{eqnarray}
\tilde{B^2} =  \frac{B^2_{in}}{g\bar{\rho_0}L_\rho} & \nonumber \\ =  {\tilde{B_1^2}} \; - & \frac{\tilde{B^2_2}}{\cosh^2[\tilde{x}_0 - \tilde{x}_{B}]} + \frac{\tilde{x} - \tilde{x}_0}{\cosh^2[\tilde{x}_0 - \tilde{x}_{\rho}]} - (\tanh [\tilde{x} - \tilde{x}_{\rho}] - \tanh [\tilde{x}_0 - \tilde{x}_{\rho}]). \nonumber \\ 
\label{normalBin}\end{eqnarray}
Then the normalised equation of motion is:
\begin{eqnarray}
\left(\frac{\partial \tilde{x}}{\partial \tilde{t}}\right)_{x_0}  
 =  \frac{\tilde{B^2}}{[1+ A^2(\frac{\partial \tilde{x}}{\partial \tilde{z}})_{x_0}^2]^2}A^2\left( \frac{\partial ^2\tilde{x}}{\partial \tilde{z}^2}\right)_{x_0} - \frac{\frac{1}{2}(\frac{\partial \tilde{B^2}}{\partial \tilde{x}})_{x_0}}{[1+ A^2(\frac{\partial \tilde{x}}{\partial \tilde{z}})_{x_0}^2]}
\label{drag1}\end{eqnarray}
and Eq.~(\ref{shoot}) for the equilibrium position of a field line becomes:
\begin{equation}
\left(\frac{\partial \tilde{x}}{\partial \tilde{z}}\right)_{x_0} = \pm\frac{1}{A}\sqrt{A^2\tilde{q}^2 + (1+A^2\tilde{q}^2)\frac{\frac{\tilde{x} - \tilde{x}_0}{\cosh^2[\tilde{x}_0 - \tilde{x}_{\rho}]} - (\tanh [\tilde{x} - \tilde{x}_{\rho}] - \tanh [\tilde{x}_0 - \tilde{x}_{\rho}]).}{{\tilde{B_1^2}} - \frac{\tilde{B^2_2}}{\cosh^2[\tilde{x}_0 - \tilde{x}_{B}]}}}
\label{shootnorm}\end{equation}
and $(\frac{\partial \tilde{x}}{\partial \tilde{z}})_{x_0} (\tilde{z}=0,1) = \tilde{q}$.  The normalised relative change in energy is:
\begin{eqnarray}
\Delta \tilde{\cal E_B}  = \frac{\cal E_B}{B_0(x_0)L} -1\;\;\;\;\;\; \;\;\;\;\; & \nonumber \\ =  \int_0^1d{\tilde z}\sqrt{1+ A^2(\frac{\partial \tilde{x}}{\partial \tilde{z}})_{x_0}^2} & \left(\sqrt{1 + \frac{\frac{\tilde{x} - \tilde{x}_0}{\cosh^2[\tilde{x}_0 - \tilde{x}_{\rho}]} - (\tanh [\tilde{x} - \tilde{x}_{\rho}] - \tanh [\tilde{x}_0 - \tilde{x}_{\rho}])}{{\tilde{B_1^2}} - \frac{\tilde{B^2_2}}{\cosh^2[\tilde{x}_0 - \tilde{x}_{B}]}}}\right) \; -\; 1. \nonumber \\
\label{normenergy}\end{eqnarray}
The weakly nonlinear behaviour (from Eq.~(\ref{ampevo})) in normalised variables is:
\begin{equation}
\frac{d{\tilde \xi}}{dt'} = {\tilde \gamma} {\tilde \xi} + {\tilde C}_4{\tilde \xi}^2
\label{weaknorm}\end{equation}
where ${\tilde \xi}\sin{{\tilde z}\pi} = {\tilde x} - {\tilde x}_0$, 
\begin{eqnarray}
{\tilde \gamma}({\tilde x}_0) =   \frac{\rho_0({\tilde x}_0)\Gamma_0^2({\tilde x}_0)L_\rho}{2g{\bar \rho_0}}   
 =  -  \left(A^2{\tilde B}_{1}^2  - \frac{A^2{\tilde B}_{2}^2}{\cosh^2{({\tilde x}_0 - {\tilde x}_{B})}}\right)
\pi^2 -\frac{\sinh{({\tilde x}_0 - {\tilde x}_{\rho})}}{\cosh^3{({\tilde x}_0 - {\tilde x}_{\rho})}}
\label{gammanorm}\end{eqnarray}
 and \begin{eqnarray}
 {\tilde C}_4 ({\tilde x}_0) = \frac{\rho_0({\tilde x}_0)C_4({\tilde x}_0)L_\rho^2}{2g{\bar \rho_0}}
= \frac{8}{3\pi}\left(\frac{3\tanh^2{({\tilde x}_0 - {\tilde x}_{\rho})}-1}{\cosh^2{({\tilde x}_0 - {\tilde x}_{\rho})}}\right)
\label{C4norm}\end{eqnarray}
Note that the nonlinear $C_4$ term drives upward motion for $({\tilde x}_0 - {\tilde x}_{\rho}) < - \tanh^{-1}(1/\sqrt{3}) \sim - 0.658$ and $({\tilde x}_0 - {\tilde x}_{\rho}) > \tanh^{-1}(1/\sqrt{3})$ and downwards motion for $\tanh^{-1}(1/\sqrt{3}) > ({\tilde x}_0 - {\tilde x}_{\rho}) > - \tanh^{-1}(1/\sqrt{3})$.

\subsubsection{Numerical Solutions for Model.}
\label{numerical}

\begin{figure}[h!]
\setlength{\unitlength}{1cm}
\begin{center}
{\includegraphics[angle=0, width=12.0cm, totalheight=11.0cm, trim=0 1cm 0 4cm, clip=true]{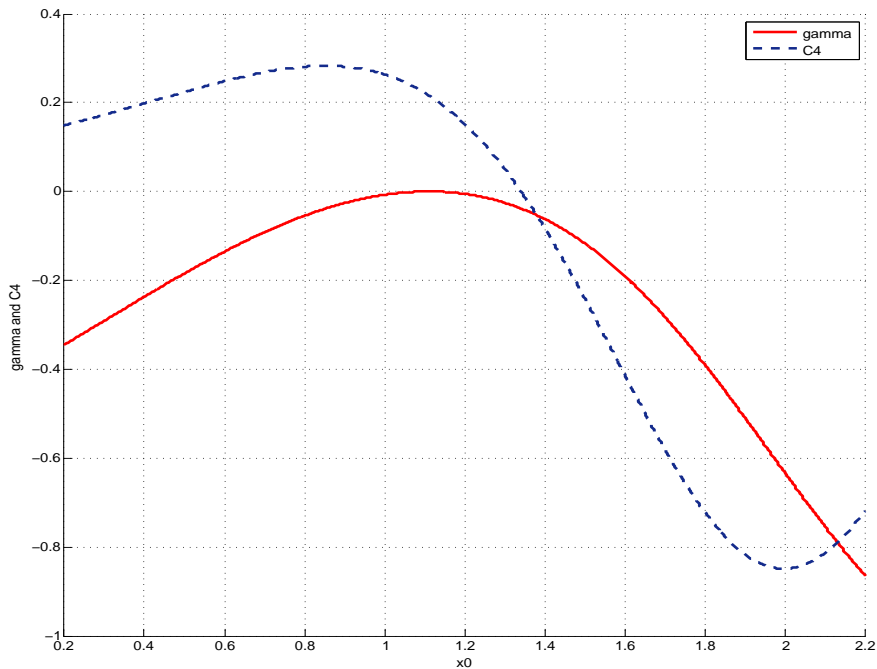}}
\caption{\textit{The growth rate ${\tilde \gamma}({\tilde x}_0)$ (red line) and the explosive nonlinearity term ${\tilde C}_4({\tilde x}_0)$ (dashed blue line) for the model equilibrium numerical cases.  The growth rate has a maximum of $1.9 \times 10^{-4}$ at ${\tilde x}_0 = {\tilde x}_{max} = 1.1118$ and is positive between $1.0945< {\tilde x}_0 < 1.1288$.
}} \label{growth}\end{center}\end{figure}

The model equilibrium is specified by five normalised parameters (ignoring ${{\bar p}_0}$ in the limit of large pressure): ${\tilde x}_{\rho}$, ${\tilde x}_{B}$, $A$, ${\tilde B}_{1}^2$ and ${\tilde B}_{2}^2$.  We investigate two cases in which we fix ${\tilde x}_{\rho}=2$, ${\tilde x}_{B}=0.8$, $A^2{\tilde B}_{1}^2 = 0.07834$ and $A^2{\tilde B}_{2}^2 = 0.04701$ and vary $A$ -- the aspect ratio of the box.  In Fig.~(\ref{growth}) we plot the growth rate ${\tilde \gamma}({\tilde x}_0)$ and the explosive nonlinearity, ${\tilde C}_4({\tilde x}_0)$, for the numerical cases -- these do not change as $A$ is changed.  Note how the system is just above marginal linear stability -- {\em i.e.} the local growth rate is slightly positive in a narrow region around ${\tilde x}_0 = 1.1118$.
\vskip 0.1 truein
The first case (CASE 1) has A=0.161604 and the second case (CASE 2) has A=0.1695.  We choose these values so that in CASE 1 all the field lines erupt without flux expulsion and in CASE 2 some of the field lines (those between ${\tilde x}_0 = 0.295$ and ${\tilde x}_0 = 0.62$) erupt into a flux expelled state.  While these cases are representative we have made no attempt yet to survey all the possible cases.  For example we have not considered cases where the field lines erupt downwards --  there could also be cases where some of the lines erupt downwards and some upwards. 

\begin{figure}[h!]
\setlength{\unitlength}{1cm}
\begin{center}
{\includegraphics[angle=0, width=12.0cm, totalheight=12.0cm, trim=0 1cm 0 4cm, clip]{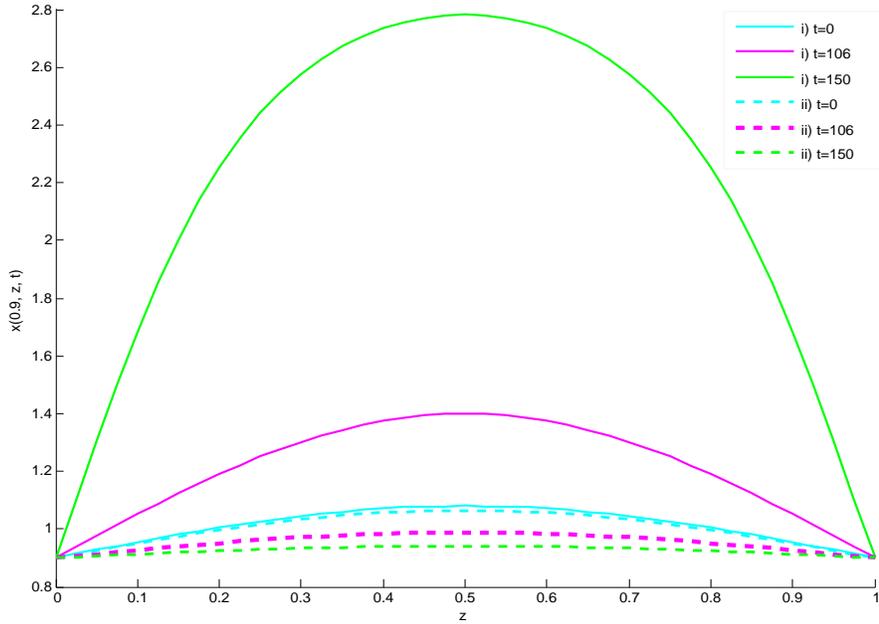}}
\vskip -1.2 truein\caption{\textit{CASE 1.  Evolution of the field line (blue $t=0$, magenta $t=106$ and green $t=150$) with $x_0 = 0.9$ from initial conditions: i) ${\tilde x}-{\tilde x}_0 = 0.18\sin{({\tilde z}\pi)}$ just above the critical amplitude for nonlinear instability (the solid lines) and; ii) ${\tilde x}-{\tilde x}_0 = 0.165\sin{({\tilde z}\pi)}$ just below the critical amplitude for instability (the dashed lines). The final state for i) is the stable equilibrium erupted state -- the saturated state(green line).  The final state for ii) is the unperturbed field line (dashed green line).}} \label{C1point9a}\end{center}\end{figure}

\begin{figure}[h!]
\setlength{\unitlength}{1cm}
\begin{center}
\vskip -0.5 truein
{\includegraphics[angle=0, width=12.0cm, totalheight=11.0cm, trim=0 1.5cm 0 4cm, clip]{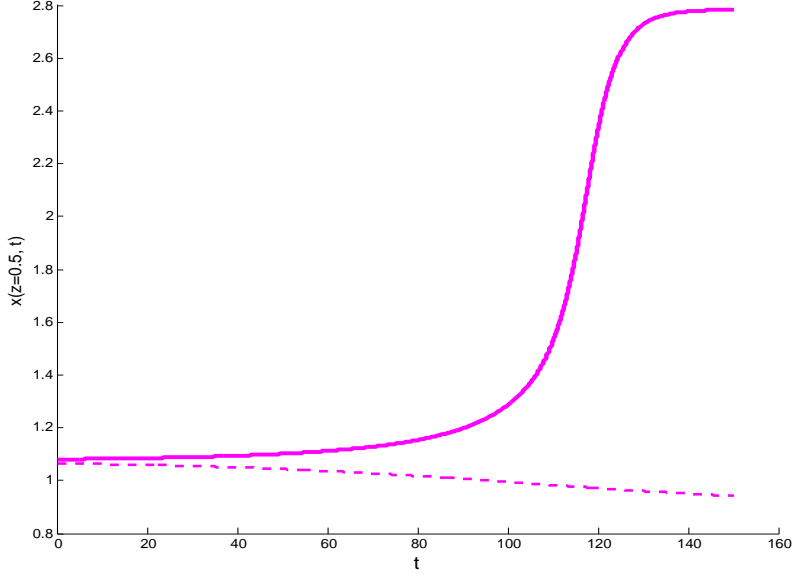}}
\vskip -1.0 truein\caption{\textit{CASE 1.  Evolution of the peak of the field line (${\tilde x}_{peak} = {\tilde x}({\tilde z}=0.5, 0.9, t)$) for: initial conditions i), (${\tilde x}({\tilde z}=0.5, 0.9, 0)= 0.18$) the solid line and;  initial conditions ii) (${\tilde x}({\tilde z}=0.5, 0.9, 0)=0.165$) - the dashed line.}} \label{C1point9b}\end{center}\end{figure}

\begin{figure}[h!]
\setlength{\unitlength}{1cm}
\begin{center}
{\includegraphics[angle=0, width=12.0cm, totalheight=11.0cm, trim=0 1.5cm 0 4cm, clip]{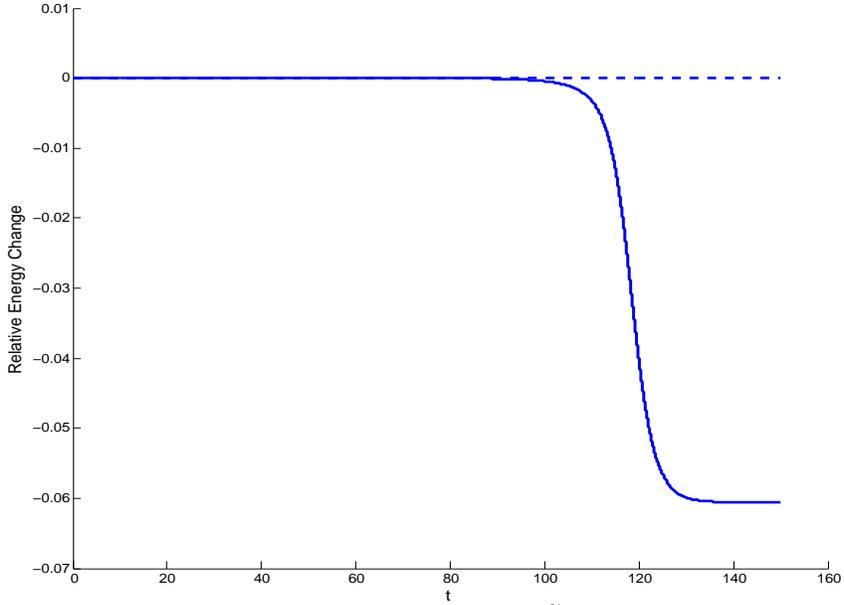}}
\vskip -1.0 truein\caption{\textit{CASE 1.  The relative energy change $\Delta \tilde{\cal E_B}$ of the ${\tilde x}_0 = 0.9$ field line as a function of time.  The initial energy in both initial conditions is almost the same $\Delta \tilde{\cal E_B}({\tilde t}) = 5.76\times 10^{-5}$.  The dashed line is ii) the below critical perturbation returning to ${\tilde x} = {\tilde x}_0$ and $\Delta \tilde{\cal E_B}({\tilde t}\rightarrow \infty) = 0$.  The solid line  shows energy released and being dissipated by drag as the above critical perturbation (i) line goes to saturation and $\Delta \tilde{\cal E_B}({\tilde t}\rightarrow \infty) = -0.06$}} \label{C1point9c}\end{center}\end{figure}

{\bf CASE 1. $A = 0.161604$}.  In this case ${\tilde B}_{1}^2 = 3$ and ${\tilde B}_{2}^2 = 1.8$ and the magnetic field energy is larger than the energy changes so that ${\tilde B}^2 > 0$ for all ${\tilde x}$ and ${\tilde x}_0$.  In Fig.~(\ref{C1point9a}) we show the evolution of ${\tilde x}(z, 0.9, t)$ ({\em i.e.} the field line with ${\tilde x}_0 = 0.9$) from initial conditions: $i)$ ${\tilde x}-{\tilde x}_0 = 0.18\sin{({\tilde z}\pi)}$ just above the critical amplitude for nonlinear instability and; $ii)$ ${\tilde x}-{\tilde x}_0 = 0.165\sin{({\tilde z}\pi)}$ just below the critical amplitude for instability.  By the time $t=150$ the field line has reached stable equilibria -- the erupted saturated state for the initial conditions $i)$ and the unperturbed state for $ii)$. In Fig.~(\ref{C1point9b}) we show the peak amplitude of the field line
(${\tilde x}({\tilde z}=0.5, 0.9, t)$) note that the exploding field line reaches saturation at a value of 
${\tilde x}({\tilde z}=0.5, 0.9, t\rightarrow \infty) = 2.78$.  We show the relative energy change $\Delta \tilde{\cal E_B}$ in Fig.~(\ref{C1point9c}).  We also solved Eq.~(\ref{shootnorm}) for the equilibrium position of the ${\tilde x}_0 = 0.9$ field  line by the shooting method (varying $q$ until we find a solution that vanishes at $z=0$ and $z=L$).  There are three equilibrium positions: the trivial ${\tilde x}(z, 0.9, t)= {\tilde x}_0 = 0.9$; the saturated state with ${\tilde x}_{peak} = {\tilde x}(0.5, 0.9, t) = 2.78$ and; the critical unstable equilibrium ${\tilde x}_{peak} = {\tilde x}(0.5, 0.9, t) = 1.07$.  The saturated state is, of course, identical with the final state of $i)$ in Fig.~(\ref{C1point9a}).  The minimum relative energy needed to excite the explosive behaviour is the energy of the critical equilibrium -- for the ${\tilde x}_0 = 0.9$ field line this is $\Delta \tilde{\cal E_B} = 5.76\times 10^{-5}$.  This is far less than the energy released going to the saturated state $\Delta \tilde{\cal E_B}({\tilde t\rightarrow\infty}) \sim 6\times 10^{-2}$
\vskip -0.9 truein

\begin{figure}[h!]
\setlength{\unitlength}{1cm}
\begin{center}
{\includegraphics[angle=0, width=16.0cm, totalheight=12.0cm, trim=0 2.5cm 0 5cm, clip]{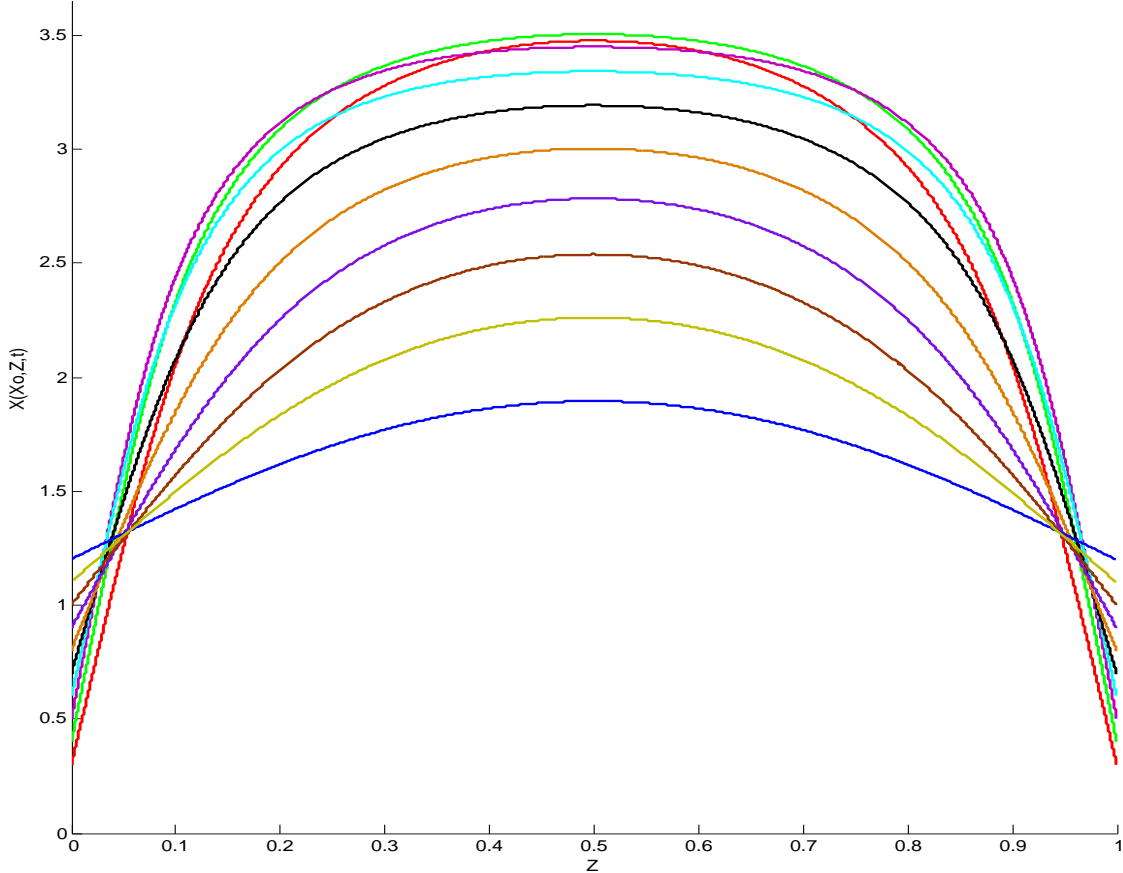}}
\caption{\textit{CASE 1.  Shapes of the saturated field lines for initial heights ${\tilde x}_0  = 0.2,\,  0.3,\,  0.4,\, 0.5,\,  0.6,\,  0.7,\,  0.8,\,  0.9,\, 1.0, \, 1.1, \,1.2$.  Note how the lower lines overtake the upper lines.}} \label{C1sat1}\end{center}\end{figure}

\begin{figure}[h!]
\setlength{\unitlength}{1cm}
\begin{center}
{\includegraphics[angle=0, width=10.0cm, totalheight=6.0cm, trim=0 4.5cm 0 5.5cm, clip]{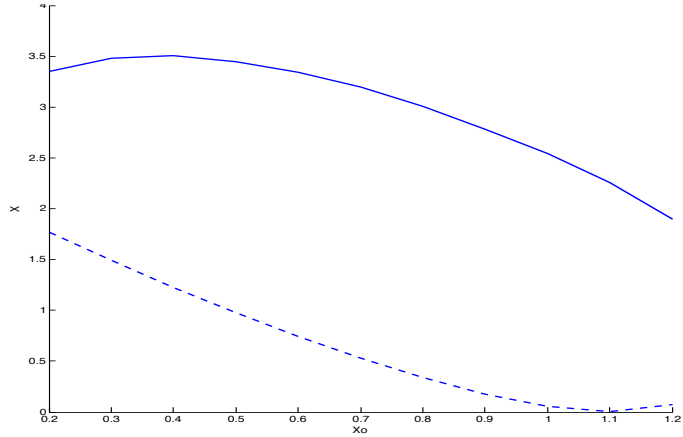}}
\vskip -0.2 truein
\caption{\textit{CASE 1.  The  peak height of the saturated equilibrium field lines ${\tilde x}_{peak}({\tilde x}_0)$ (the solid line) and the peak displacement of the critical equilibrium field lines (the dashed line).  Note that in the linearly unstable region around $x_0=1.1$ the critical displacement is zero.  Again note that from $x_0 = 0.4$ to $x_0 = 1.2$ the peak height is a decreasing function of initial height -- {\em i.e.} the lines overtake.}} \label{C1sat/crit2}\end{center}\end{figure}

\begin{figure}[h!!]
\setlength{\unitlength}{1cm}
\begin{center}
{\includegraphics[angle=0, width=12.0cm, totalheight=7.5cm, trim=0 4.5cm 0 5.5cm,  clip]{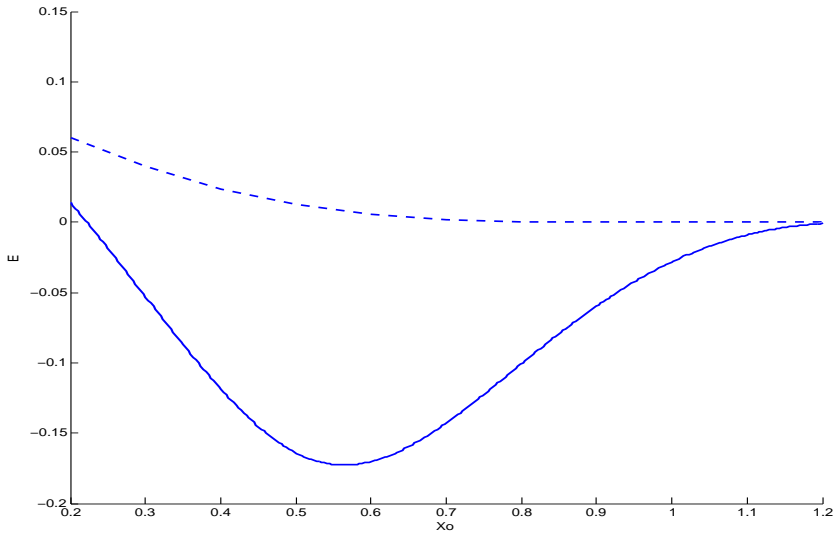}}
\vskip -0.2 truein
\caption{\textit{CASE 1.  The relative energy $\Delta \tilde{\cal E_B} = E$ of saturated equilibrium field lines (the solid line) and relative energy $\Delta \tilde{\cal E_B} = E$ of the critical equilibrium field lines (the dashed line).  The energy of the critical equilibrium is the minimum energy to excite explosive motion.  Note that below $x_0\sim 0.23$ the saturated equilibrium is a higher energy than the initial state.  From $x_0\sim 0.23$ to $x_0\sim 1.2$ the saturated state has a lower energy than the initial state and therefore energy can be released.  The largest relative energy change is the $x_0\sim 0.57$ field line that can release about $17\%$ of its initial magnetic energy from a critical perturbation that is $1\%$ of its initial magnetic energy.}}  
\label{E1sat/crit3}\end{center}\end{figure}

\newpage
Using the shooting method and Eq.~(\ref{shootnorm}) we have computed the saturated states and critical equilibrium states for the field lines from ${\tilde x}_0 = 0.2$ to ${\tilde x}_0 = 1.2$ (in steps of $\Delta{\tilde x}_0 = 0.1$).  Below $x_0 \sim 0.2$ and above $x_0\sim 1.25$ we found no equilibrium states other than the initial state.  The saturated field line shapes are shown in Fig.~(\ref{C1sat1}).  Note how the lower lines overtake the upper lines.  In Fig.~(\ref{C1sat/crit2}) we plot the saturated height and critical displacement as a function of the original field line position {\em i.e.} ${\tilde x}_{peak}({\tilde x}_0)$ and in Fig.~(\ref{E1sat/crit3}) we plot the relative energy of the saturated state and the critical energy to excite the nonlinear explosive motion.  
\vskip -0.1 truein
\vskip -0.3 truein
\vskip -0.9 truein

\newpage

\newpage

\begin{figure}[h!]
\setlength{\unitlength}{1cm}
\begin{center}
{\includegraphics[angle=0, width=10.0cm, totalheight=6.0cm, trim=0 4.5cm 0 5.5cm,  clip]{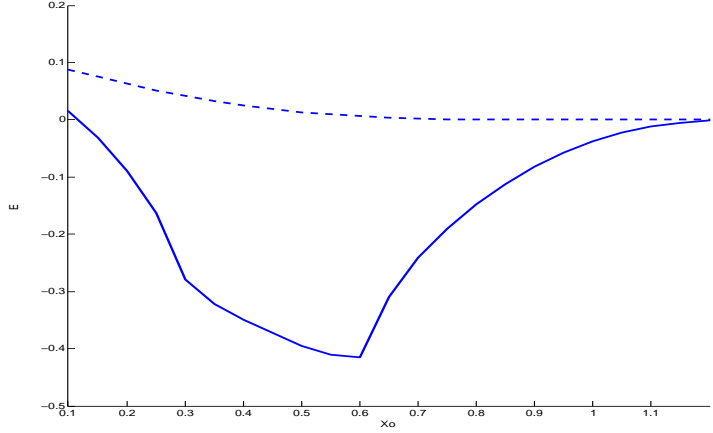}}
\vskip -0.2 truein
\caption{\textit{CASE 2.  The relative energy $\Delta \tilde{\cal E_B} = E$ of saturated equilibrium field lines (the solid line) and critical relative energy $\Delta \tilde{\cal E_B} = E$ of the critical equilibrium field lines (the dashed line).  Note that below $x_0\sim 0.13$ the saturated equilibrium is a higher energy than the initial state.  From $x_0\sim 0.13$ to $x_0\sim 1.2$ the saturated state has a lower energy than the initial state and therefore energy can be released.  From $x_0\sim 0.295$ to $x_0\sim 0.62$ the saturated field lines have a rectangular shape with $B_{in} =0$ for $x_{peak} = x_{zero}$.  The largest relative energy change is the $x_0\sim 0.6$ field line that can release about $41\%$ of its initial magnetic energy from a critical perturbation that is only $0.6\%$ of its initial magnetic energy.}}  
\label{C2Esat}\end{center}\end{figure}

{\bf CASE 2. $A = 0.1695$}.  In this case ${\tilde B}_{1}^2 = 2.72701$ and ${\tilde B}_{2}^2 = 1.63620$.  The minimum value of ${\tilde B}^2$ (or equivalently $B^2_{in}$) for a given ${\tilde x}_0$ is at ${\tilde x} = 2{\tilde x}_{\rho} - {\tilde x}_{0} = 4 - {\tilde x}_{0}$.  Field lines between $x_0 = 0.295$ and $x_0 = 0.62$ have a minimum value of ${\tilde B}^2$  that is less than zero.  Thus this case has a region of flux tubes where  {\em flux expulsion} takes place as treated in Section~(\ref{Expulsion}).  All the field lines that have a minimum ${\tilde B}^2$ less than zero ({\em i.e. } field lines with $0.295 < x_0 < 0.62$) minimise their energy $\tilde{\cal E_B}$ by taking the limiting rectangular shape of Section~(\ref{Expulsion}).  Thus for these field lines we evaluate
$\Delta\tilde{\cal E_B}$ using (the appropriately normalised) Eq.~(\ref{energymin}) for all others we use Eq.~(\ref{normenergy}).  The critical energy and the saturated energy are plotted in Fig.~(\ref{C2Esat})

\newpage

%
%
%
%
%
\section{Discussion and Conclusion.}
\label{discussion}

The calculations presented in this paper support a model of explosive release of energy  in magnetised atmospheres by the destabilisation of multiple metastable flux tubes.   This eruption model is far from complete; indeed a number of questions remain.  Nonetheless some results are clear.
In Section~(\ref{fluxtube}) we demonstrate the metastability of isolated thin elliptical flux tubes in a magnetised atmosphere.  We show that tubes can erupt on Alfv\'{e}nic timescales  when they cross the linear stability boundary, or when they are displaced by an amplitude greater than the critical amplitude.    With viscous (or drag) dissipation the flux tubes will relax to finitely displaced (saturated) equilibrium states releasing a significant fraction of their stored energy.  The energy needed to destabilise all the metastable tubes is considerably less than the energy released (see Figures~(\ref{E1sat/crit3}) and (\ref{C2Esat})).  In some high pressure cases the saturated equilibrium state is singular and the flux tube swells to infinite thickness thus reducing the field in the tube to zero -- see Section~(\ref{Expulsion}).  We have also shown (see Section~(\ref{pert})) that the weakly nonlinear behaviour near marginal stability yields growth in a narrow unstable region with erupting fingers pushing into and progressively destabilising the metastable region -- a process we have called {\em detonation}.  More detail of this mechanism is given in \cite{fong2, fong, CWHF, wilcow}.  We have conjectured that these fingers evolve into flux tube eruptions.  In the rest of this discussion section we address unresolved issues for our eruption model qualitatively.

The size, number and shape of the flux tubes in an eruption must depend to some extent on the noise that creates the perturbation.  We distinguish between slow evolution of the equilibrium and {\em noise} perturbations that temporarily move the system out of equilibrium.  A large perturbation of a linearly stable plasma could trigger energy release but, at least in the early stage of eruption, the shape of the perturbation must determine the tubes that participate.  However noise levels are usually small in systems of interest.  Thus large perturbations are rare and would themselves require an explanation.  In fusion experiments the background {\em drift wave} turbulence provides a constant source of weak low frequency noise with density perturbations of a few percent at most.  The noise in the solar corona comes from the convective motions that slowly perturb the foot points of the field lines.  It seems likely, therefore, that eruptions begin in a region that is very close to being marginally stable with perturbations (some part of the noise spectrum) that are close to the most unstable linear perturbations.  The weakly nonlinear dynamics of marginally unstable atmospheres (see Section~(\ref{pert})) shows that the dynamics evolves into a number of interacting explosively growing fingers.  We argue that this is the beginning of the eruption of elliptical flux tubes -- but not necessarily isolated elliptical flux tubes. 

We have assumed that the erupting flux tubes are strongly elliptical in shape {\em i.e.} $\delta_1 \ll \delta_2$ in Figure~(\ref{tube}).  The shape of the eigen-mode in the linear regime has $\delta_1 \sim {\cal{O}}(\epsilon^2)L_\gamma$ and $\delta_2 \sim {\cal{O}}(\epsilon)L_\gamma$ where $\epsilon \sim \gamma/\Gamma_A\ll 1$ and $\gamma$ is the growth rate, $\Gamma_A$ the Alfven frequency and the typical equilibrium scale length is $L_\gamma$ (see Subsection~(\ref{linear})).  The evolution in the weakly nonlinear regime depends on the dissipation -- here the narrowing of the width of fingers in $y_0$ depends on viscosity (in fusion finite larmor radius physics controls the narrowing, see \cite{fong, fong2}).  If the resistivity is larger than the viscosity (in a small magnetic Prandtl number plasma) the eruption would be very different as the plasma would disconnect from the field.  We will continue to assume that the resistive diffusion across the flux tube is slower than the eruption time.  As a tube erupts it will push the tubes in front of it causing some of them to become destabilised, it will also drag up tubes from below.  This transfer of energy to metastable tubes is the {\em detonation} mechanism -- we do not have a good understanding of its efficiency.  If the detonation by each tube is efficient then we expect the whole height of the metastable region to erupt -- in Subsection~(\ref{model}) -- this would lead to $\delta_2 \sim L_\rho$ see Figure~(\ref{C1sat1}).  The field lines surrounding the flux tube (coloured blue in Figure~(\ref{tube})) are bent sideways (in the $y$ direction).  They will tend to flatten the tube with a force of order $B_0^2\delta_1/L^2$ in the $y$ direction.  This is however small compared to the forces in the $x$ direction and therefore we expect it to  make little difference to the final state.

\begin{figure}[h!]
\setlength{\unitlength}{1cm}
\begin{center}
{\includegraphics[angle=0, width=10.0cm, totalheight=6.0cm, trim=0 2cm 0 1.5cm,  clip]{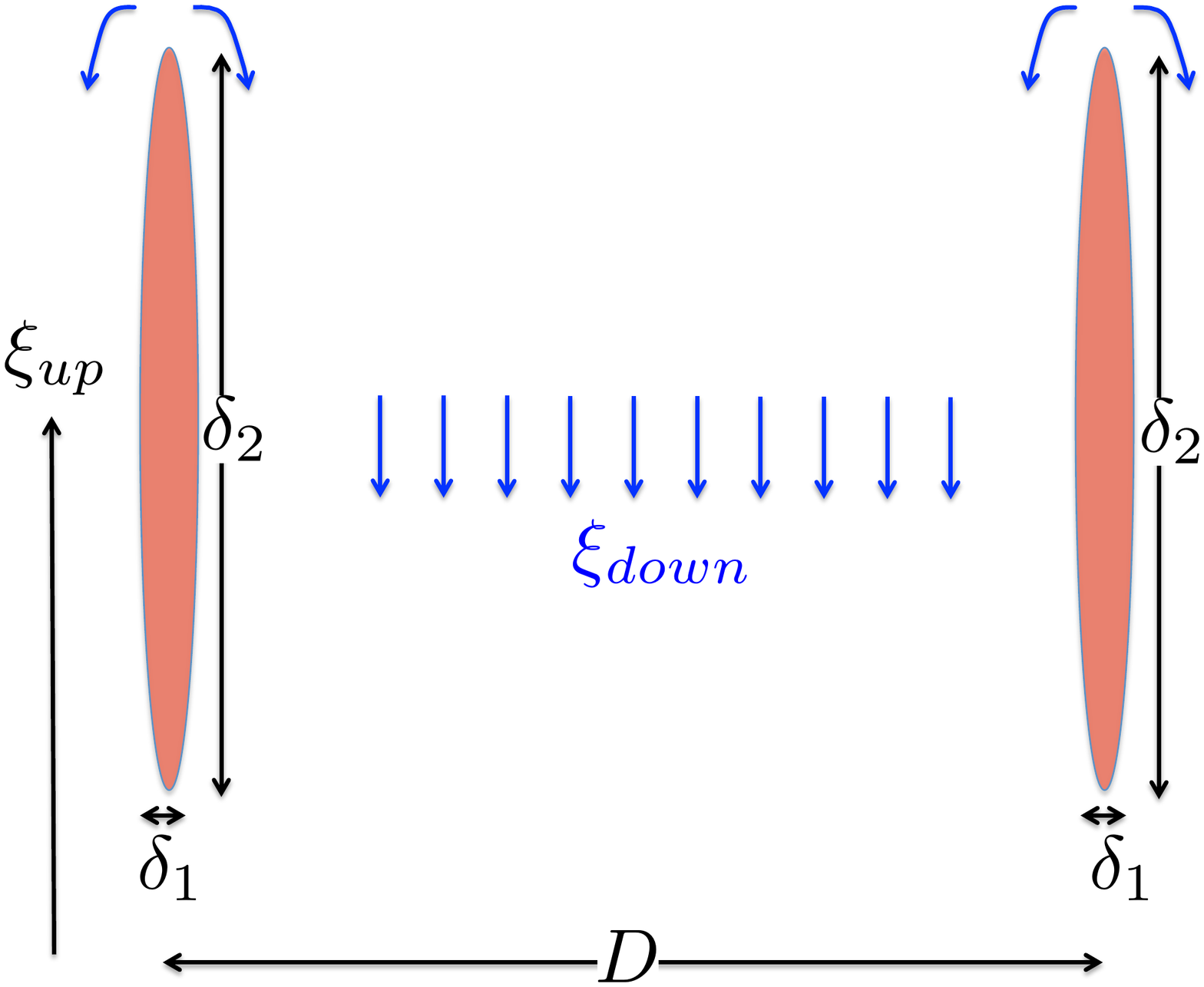}}
\caption{\textit{Cross section of two (red) tubes spaced by $D$ in the $y$ direction.  The tubes have width $\delta_1$ and vertical height $\delta_2$ and have been displaced an average distance of $\xi_{up}$ upwards.  The motion of the surrounding plasma is shown by blue arrows.  Field lines between the tubes are displaced downwards an average distance of $\xi_{down}$.}}  
\label{Multi-tubes}\end{center}\end{figure}

If detonation was completely efficient then we would expect the final state to be the lowest possible energy state.  Such a state must pack in as many erupted tubes as possible.  Let us now estimate the ($y$) distance between tubes ($D$) which maximises the energy release.  In Figure~(\ref{Multi-tubes}) we illustrate two tubes spaced by $D$ and displaced upwards by an average distance of $\xi_{up}$.  Typically $\xi_{up}$ is of order $L_\rho$ the gravitational scale height and as discussed above with efficient detonation we expect $\delta_2 \sim L_\rho$.  The plasma between the tubes must be displaced down to make room for the flux tube rising.  The plasma is compressible but we will estimate the downwards displacement from rough incompressibility -- $\delta_1\delta_2 \sim \xi_{up}\delta_1\sim \xi_{down} D$.  We can estimate the energy needed to drive the downward motion as a fraction $A_1 <1$ of the field line bending energy {\em i.e.} $A_1(\xi_{down} B_0/L)^2D\delta_2 L = A_1 (\xi_{up} B_0/L)^2(\delta_1/D)^2D\delta_2 L$.  The energy available from the upward going flux tube is some fraction $A_2 <1$ of the gravitational energy {\em i.e.} $A_2 (\rho_0g\xi_{up})(\delta_1\delta_2 L)$.  The sideways motion of the field lines next to the flux tube gives another stabilising energy $(\delta_1 B_0/L)^2(\delta_1\delta_2 L)$ which is always small for elliptical tubes.  We release energy if the energy from the upwards moving flux tube exceeds the energy from the downwards motion {\em i.e. if} 
\begin{equation}
D > \delta_1\frac{A_1}{A_2}(\frac{B_0^2L_\rho}{\rho_0 gL^2})(\frac{\xi_{up}}{L_\rho})
\end{equation}

Note the factor $(B_0^2L_\rho)/(\rho_0 gL^2)$ is of order one for a profile near marginal stability. Therefore we can release energy if we make $D$ bigger than $\delta_1$ by a finite factor.  The total energy release is then a finite fraction of the energy available if the magnetic field were absent -- the gravitational energy $\sim \rho_0 g L_\rho V_0$ where $V_0$ is the volume of the metastable region.

Energy in the eruption is dissipated by viscosity or aerodynamic drag in our model.  This would yield simple ion heating.  If the motion does approach the sound speed we would expect the formation of shocks and possible acceleration of particles to non thermal energies.  The elliptical flux tubes in the saturated state have two current sheets, one on each side of the flux tube.  These current sheets have opposite sign.  Even if resistive diffusion is negligible during the eruption it could act on a longer timescale on the saturated state reconnecting the field lines releasing more energy as heat.  Such resistive diffusion would smooth out the magnetic field in the final state.   We have ignored many secondary effects of the eruption; for example we have not considered the possibility of secondary instabilities driven by the gradients across the flux tube.  

The discussion in this section is speculative.  Clearly high resolution simulations that can follow the eruption to saturation would help resolve the many unanswered questions. Future work will pursue these questions and the extension of these ideas to more complex equilibria.

\newpage

\section*{Acknowledgements} Steve Cowley would like to acknowledge discussions on the issues presented in this paper with Omar Hurricane, Bryan Fong, Alex Schekochihin, Chris Ham, Jack Connor and Felix Parra.  This project has received funding from the European Union's Horizon 2020 research and innovation programme under grant agreement number 633053 and from the RCUK Energy Programme [grant number EP/I501045]. Sophia Henneberg  is funded by the German Academic Exchange Service (DAAD - Stipendium fŸr Doktoranden). Howard Wilson is a Royal Society Wolfson Research Merit Award Holder.  To obtain further information on the data and models underlying this paper please contact PublicationsManager@ccfe.ac.uk. The views and opinions expressed herein do not necessarily reflect those of the European Commission.

\newpage
\begin{appendix}
\section{Threshold for Instability in Slab.}
\label{threshold}
In this Appendix we outline the proof that infinitesimally above the threshold for instability (for the slab equilibria) the unstable perturbations have wave numbers in the $y$ direction that asymptote to infinity ({\em i.e.} $k_y\rightarrow\infty$).  The proof is a simple extension for our chosen boundary conditions of the proof in \cite{Zweibel} which followed \cite{newcomb} and \cite{gilman}.  From Eqs.~(\ref{moment}), (\ref{magnetic}), (\ref{density}) and $p = \rho T_0(x_0)/m$ the equation of motion for a linear displacement of the plasma (dropping viscosity) becomes:
 \begin{eqnarray}
 \rho_0\frac{\partial^2 \bxi}{\partial t^2} = {\bf F}(\bxi)& = &\grad\left[ (p_0 + B_0^2)\grad\cdot\bxi - \xi_x\rho_0g - B_0^2\frac{\partial \xi_z}{\partial z}\right]
\nonumber \\ & + & B_0^2\frac{\partial^2 \bxi}{\partial z^2} - B_0\left(\frac{\partial  }{\partial z}\grad\cdot\bxi\right){\bf B_0} + (\xi_x g\frac{d\rho_0}{dx} + \rho_0 g \grad\cdot\bxi){\xx}.
\label{forceMHD}
\end{eqnarray}
The line tied boundary conditions with $\delta\rho=\delta p =0$ at $z=0,L$ yield the boundary conditions on $\bxi$:
\begin{equation}
\xi_x=\xi_y = \frac{\partial  \xi_z}{\partial z} =0 \rightarrow \grad\cdot\bxi =0 \;\;\; at \;\;\; z=0,\ L.
\label{bcs}\end{equation}
With these boundary conditions ${\bf F}(\bxi)$ is self-adjoint {\em i.e.} $\int d^3r (\boldeta\cdot{\bf F}(\bxi)) = \int d^3r (\bxi\cdot{\bf F}(\boldeta))$ when both $\bxi$ and $\boldeta$ satisfy the boundary conditions of Eq.~(\ref{bcs}).  Therefore there exists an energy principle - \cite{bernstein}.  The potential energy is:
 \begin{eqnarray}
&&\delta W(\bxi, \bxi) = \frac{1}{2}\int d^3r (\bxi\cdot{\bf F}(\bxi)) \nonumber \\  
 & = &\frac{1}{2}\int d^3r \left[ (p_0 + B_0^2)(\grad\cdot\bxi)^2 - 2(\grad\cdot\bxi)(B_0^2\frac{\partial \xi_z}{\partial z} + \xi_x\rho_0g) + B_0^2(\frac{\partial \bxi}{\partial z})^2 -\xi_x^2 g\frac{d\rho_0}{dx}\right]. \nonumber \\ 
\label{deltaW}
\end{eqnarray}
The plasma is linearly unstable if and only if there is a displacement for which $\delta W <0$.  Suppose we take a displacement with $y$ wavenumber $k_y$  of the general form
\begin{equation}
\bxi(k_y) = \left({\hat\xi}_x(x,z)\cos{(k_y y + \phi)}, \frac{1}{k_y}{\hat\xi}_y(x,z)\sin{(k_y y+\phi)}, {\hat\xi}_z(x,z)\cos{(k_y y +\phi)}\right)
\label{minimum}\end{equation}
where $\phi$ is an arbitrary constant phase.  Further suppose that the functions ${\hat\xi}_x(x,z)$, ${\hat\xi}_y(x,z)$ and ${\hat\xi}_z(x,z)$ minimise $\delta W$ for fixed $k_y$.  In general the solution of the Euler Lagrange equations yield nonzero ${\hat\xi}_y$ and therefore nonzero $\frac{\partial {\hat\xi}_y}{\partial z}$.  Let $\bxi(k_y')$ be given by the same expression as $\bxi(k_y)$ in Eq.~(\ref{minimum}) with $k_y\rightarrow k_y'$.  Then from Eq.~(\ref{deltaW}) we obtain:
\begin{equation}
\delta W(\bxi(k_y), \bxi(k_y)) -  \delta W(\bxi(k_y'), \bxi(k_y')) = \frac{1}{4}\int d^3r [(\frac{1}{k_y^2} - \frac{1}{k_y'^2})B_0^2(\frac{\partial {\hat\xi}_y}{\partial z})^2].
\end{equation}
Therefore if $k_y < k_y'$ then $\delta W(\bxi(k_y'), \bxi(k_y')) < \delta W(\bxi(k_y), \bxi(k_y))$.  Clearly we can always decrease $\delta W$ by increasing $k_y$.  Thus the minimum $\delta W$ (which at the marginal stability threshold must have the value zero) must result from a displacement with $k_y\rightarrow\infty$.  This proves the assertion that just above the marginal threshold the unstable perturbations have $k_y\rightarrow\infty$.

\section{Erupting Flux Tubes in a General Equilibrium}
\label{general}
\begin{figure}[h!]
\setlength{\unitlength}{1cm}
\begin{center}
{\includegraphics[angle=0, width=14.0cm, totalheight=8.5cm, trim=2.0cm 1.5cm 2.0cm 1.5cm,  clip]{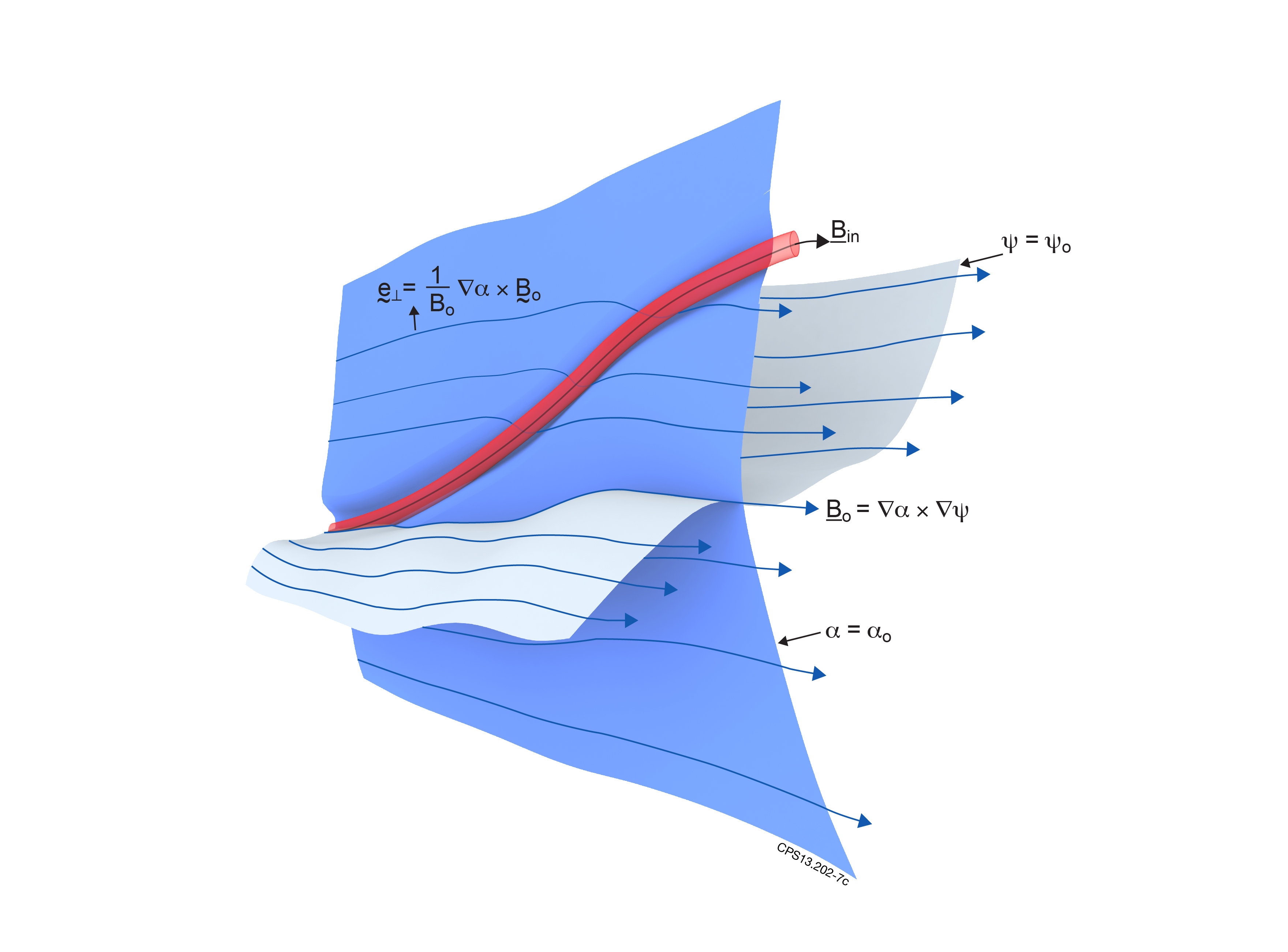}}
\vskip 0.2 truein
\caption{\textit{An elliptical (red) flux tube erupting along the surface $\alpha = \alpha_0$.  The external (blue) field lines are only slightly perturbed.  The central field line of the tube comes from the surface $\psi = \psi_0$.  The equations for this field line in Clebsch coordinates are $\psi = \psi(\psi_0, l, t)\;\;and \;\;\; \alpha = \alpha_0$.}}  
\label{gentube}\end{center}\end{figure}

In this section we generalise the treatment of Section~(\ref{fluxtube}) to the dynamics of single isolated flux tube in a general stationary magnetic equilibrium.   We shall assume that the tube is moving somewhat slower than the sound speed since we are interested in the behaviour near marginal stability and the saturated states of the flux tube.  Consider a field aligned tube of plasma that is displaced through the plasma, see Fig.~(\ref{gentube}).  The field inside the tube is denoted ${\bf B}_{in}$ and the field outside ${\bf B}_{out}$ -- see Fig.~(\ref{gentube}).   The tube has an elliptical cross section, elongated in the direction of motion and narrower across ($\delta_1 \ll\delta_2 $), see Fig.~(\ref{gentube}).    Again the exact cross sectional shape of the tube is not important here -- just that it is narrow enough that the perturbation of the surrounding field is unimportant and that it is considerably elongated in the direction of motion.  

As the erupting tube moves it must follow a surface, ${\cal S}$, which is tangent to both the tube and the surrounding field lines, see Fig.~(\ref{gentube}) .  We shall again assume that the surrounding field is largely unperturbed.  We can therefore take the surface ${\cal S}$ to be a surface $\alpha = constant=\alpha_0$ where $\alpha$ is a Clebsch potential  of the unperturbed field {\em i.e.} $\bb = \grad \psi \times \grad \alpha$.  Clearly the surface ${\cal S}$ twists (see Fig.~(\ref{gentube}))  -- the local twist is a measure of the {\em local shear}.   The choice of Clebsch potentials is not unique or always single valued -- we will assume that in this case it can be single valued over the domain of interest.  We can change $\alpha$ by the transformation $\alpha\rightarrow \alpha + f(\psi)$ with arbitrary $f(\psi)$ without changing $\bb$.  It is not {\em a priori} obvious how to choose ${\cal S}\equiv\alpha$ -- {\em i.e.} which Clebsch surface the flux tube chooses to erupt along.  Indeed it is likely to be determined by the dynamics. We will derive equations for a general choice of ${\cal S}\equiv\alpha$. 

To describe the position of the flux tube we use flux coordinates of the unperturbed field: $\psi$, $\alpha$ and $l$ where $l$ is a measure of distance along the field line such that $\bb\cdot\grad l =  \grad\psi\times\grad\alpha\cdot\grad l\neq 0$ (but otherwise unspecified).
Then a field line inside the erupting tube can be described by, the value of $\alpha $ corresponding to ${\cal S}$, the value of $\psi$ and $l$.  Specifically the equation for the field line is:
\begin{equation}
\alpha = \; constant\; = \alpha_0, \;\; \psi = \psi_0 + {\tilde \psi}(l, \psi_0)
\label{appsitwid}\end{equation}
where $\psi_0$ and $\alpha_0$ are the flux coordinates of the field line before it erupted.  Since we assume that no reconnection has taken place ${\tilde \psi}(l, \psi_0)\rightarrow 0$ as $|l|\rightarrow\infty$.  Note that we do not consider any $\alpha$ dependence inside the tube or any small motions across the tube -- thus the line is supposed to lie on ${\cal S}$, the constant $\alpha$ surface.  The field in the tube must be everywhere perpendicular to $\grad \alpha$ so that it lies in the surface ${\cal S}$.  Thus we can write:
\begin{equation}
{\bf B}_{in} = K[ \grad \psi \times \grad \alpha + h \grad \alpha \times \grad l ] = K[ \bb + h \grad \alpha \times \grad l ] 
\label{apBin1}\end{equation}
where $K$ and $h$ are functions of $l$, $\alpha_0$ and $\psi_0$ (either explicitly or implicitly through dependence on $\psi$) to be found.  We will suppress the dependence on $\alpha_0$ since it is not needed for the rest of the derivation.  Using Eqs.~(\ref{appsitwid}) and (\ref{apBin1}) we obtain:
\begin{equation}
\frac{{\bf B}_{in} \cdot\grad\psi}{{\bf B}_{in} \cdot\grad l} =  \left(\frac{\partial \psi}{\partial l}\right)_{\psi_0} =   \left(\frac{\partial {\tilde\psi}}{\partial l}\right)_{\psi_0} =h(l, \psi_0).
\label{aplinetwid}\end{equation}
It is convenient to write ${\bf B}_{in}$ in terms of orthogonal vectors as:
\begin{equation}
{\bf B}_{in}  = K\left[ \bb \left(1 + s\left(\frac{\partial {\tilde\psi}}{\partial l}\right)_{\psi_0}\right) + {\bf e}_\perp u\left(\frac{\partial {\tilde\psi}}{\partial l}\right)_{\psi_0}  \right]
\label{apBin2}\end{equation}
 where:
\begin{eqnarray}
{\bf e}_\perp =  \frac{1}{B_0} \grad \alpha \times \bb , \;\;\;\;\;
u  = \frac{1}{B_0}{\bb}\cdot\grad l , \;\;\;\; 
s =   - \frac{1}{B_0}{\bf e}_\perp\cdot\grad l.
\end{eqnarray}

The force (per unit volume) on the plasma is: 
\begin{equation}
{\bf F} = -\grad [p + \frac{B^2}{2}] + {\bf B}\cdot\grad {\bf B} - \rho \grad\phi
\label{apforce}\end{equation}
where $\phi$ is the gravitational potential.  The force across the narrow tube (in the $\grad\alpha$ direction) is formally large, ${\cal O} (p/\delta_1)$, and must cancel to this order {\em i.e.}
\begin{equation}
{\bf F}\cdot\grad\alpha \sim - |\grad\alpha|^2\frac{\partial}{\partial\alpha} [p + \frac{B^2}{2}] =0.
\label{apalphaforce}\end{equation}
Thus on the slow evolution time:
\begin{equation}
p_{in} + \frac{B_{in}^2}{2} = p_{out} + \frac{B_{out}^2}{2},
\label{aptotal}
\end{equation}
where "$in$" refers to inside the tube and "$out$" refers to just outside the tube (at the same $\psi$ and $l$ along the tube -- see Fig.~(\ref{gentube})).  We will assume that the field and pressure outside the tube are unperturbed so that: 
\begin{equation}
 p_{out} = p_0(\psi, l) \;\;\;\; and \;\;\;  B_{out} = B_0(\psi, l)
\label{appressure_out}\end{equation}
are known (note again that we are suppressing dependence on $\alpha$ since everything is on the surface $\alpha = \alpha_0$). The total pressure forces at a point on the tube are thus identical to the total pressure forces on the plasma it replaced.  

We will assume for simplicity that both the unperturbed and perturbed field that the temperature is constant along the field line.  Parallel thermal conduction is often fast enough that this is true.  Then without loss of generality we can choose constant $\psi$ surfaces to coincide with constant temperature surfaces of the unperturbed state -- {\em i.e.} $T_0 = T_0(\psi)$.  Also, since we are interested in stable  displaced equilibrium states of the flux tube and slow drag dominated eruptions from an unstable state we set ${\bf F\cdot B} = 0$ for both the unperturbed and perturbed field lines.
Thus for the unperturbed field lines:
\begin{equation}
{\bf B}_{0}\cdot\grad p_{0} = \frac{T_0(\psi)}{m}{\bf B}_{0}\cdot\grad \rho_{0} = -\rho_0{\bf B}_{0}\cdot\grad\phi \;\;\rightarrow \;\; p_{0}(\psi, l) = {\bar p}_0(\psi)e^{-\frac{m\phi(\psi, l)}{T_0(\psi)}}
\label{appressure_0}\end{equation}
where $p_0 = \frac{T_0(\psi)}{m}\rho_0$.  In the flux tube we assume $T = T_0(\psi_0)$ and therefore
\begin{equation}
{\bf B}_{in}\cdot\grad p_{in} = \frac{T_0(\psi_0)}{m}{\bf B}_{in}\cdot\grad \rho_{in} = -\rho_{in}{\bf B}_{in}\cdot\grad\phi \;\;\rightarrow \;\; p_{in}(\psi, l, \psi_0) = {\bar p}_0(\psi_0)e^{-\frac{m\phi(\psi, l)}{T_0(\psi_0)}}\label{appressure_in}\end{equation}
since the field line is connected to the $\psi_0$ surface.  The approximation of parallel equilibrium for the erupting tube is correct when motions are slow compared to the sound transit time (the sound transit time is $l_\parallel/C_s$ where $l_\parallel$ is the typical scale along the field line of the variation of the displacement  and $C_s = \sqrt{ p/\rho}$ is the sound speed).  From Eqs~(\ref{aptotal}),(\ref{appressure_out}), (\ref{apforce}) and (\ref{appressure_in}) we obtain:
\begin{eqnarray}
{B}_{in}^2  =   B_0^2(\psi, l) + 2[p_0(\psi, l) - p_{in}(\psi, l, \psi_0)]
\label{B_inscalar}\end{eqnarray}
note that:
\begin{eqnarray}
\frac{1}{2}{\bf e}_\perp\cdot\grad_{\psi_0}{B}_{in}^2  =   {\bf e}_\perp\cdot\left( \bb\cdot\grad\bb  - (\rho_{0} - \rho_{in}) \grad\phi \right)
\label{B_ingrad}\end{eqnarray}
Here the gradient -- $\grad_{\psi_0}$ -- is taken at constant $\psi_0$ (as indicated), $\rho_{in}= \frac{m}{T_0(\psi_0)}p_{in}(\psi, l, \psi_0)$, $\rho_{0}= \frac{m}{T_0(\psi)}p_{0}(\psi, l)$ and $\bb\cdot\grad\bb$ is evaluated at $\psi$ and $l$.  Using Eq.~(\ref{apBin2}) we obtain:
\begin{eqnarray}
{\bf B}_{in} & = &  B_{in} { \;\bf b}_{in} \nonumber \\\nonumber \\
& = & \frac{B_{in}}{\sqrt{{\left(1 + s\left(\frac{\partial {\tilde\psi}}{\partial l}\right)_{\psi_0}\right)^2 + w^2\left(\frac{\partial {\tilde\psi}}{\partial l}\right)_{\psi_0}^2}}}\left[ \left(1 + s\left(\frac{\partial {\tilde\psi}}{\partial l}\right)_{\psi_0}\right)\frac{{\bf B}_0}{B_0} +  \frac{u}{B_0}\left(\frac{\partial {\tilde\psi}}{\partial l}\right)_{\psi_0}{\bf e}_\perp  \right] \nonumber \\ \nonumber \\
& = & a{\bf B}_0 + c{\bf e}_\perp
\label{apB_intot}\end{eqnarray}
Where ${\bf b}_{in}$ is a unit vector in the direction of ${\bf B}_{in}$.  The two coefficients, $a= a(\psi, \psi_0, \frac{\partial {\tilde\psi}}{\partial l},  l)$ and $c= c(\psi, \psi_0, \frac{\partial {\tilde\psi}}{\partial l},  l)$, are defined by Eq.~(\ref{apB_intot}) and $w^2 = \frac{u^2|{\bf e}_\perp|^2}{B_0^2}$.

The force in the tube in the direction of motion (${\bf e}_\perp$) is:
\begin{eqnarray}
{\bf F}\cdot { {\bf e}_\perp } = F_\perp & = & [{\bf B}_{in}\cdot\grad {\bf B}_{in} - \grad (p_{in} + \frac{B_{in}^2}{2}) - \rho_{in} \grad\phi ]\cdot { {\bf e}_\perp } \nonumber \\ & = &[{\bf B}_{in}\cdot\grad {\bf B}_{in} - {\bf B}_{0}\cdot\grad {\bf B}_{0} + (\rho_{0}  - \rho_{in}) \grad\phi]\cdot { {\bf e}_\perp }\nonumber \\ & = &[{\bf B}_{in}\cdot\grad {\bf B}_{in} - \frac{1}{2}\grad_{\psi_0}{B}_{in}^2]  \cdot { {\bf e}_\perp }
\label{apcurve}
\end{eqnarray}
where we have used Eqs~(\ref{aptotal}), (\ref{appressure_out}) and (\ref{B_ingrad}) and equilibrium force balance $\grad [p_0 + \frac{B_0^2}{2}] = {\bf B}_0\cdot\grad {\bf B}_0 - \rho_{0} \grad\phi$.  We have dropped corrections of order $\delta_1/\delta_2$.  Note that the force in the second expression of Eq.~(\ref{apcurve}) is a generalisation of {\em Archimedes principle} with the first two terms being the difference of the curvature forces inside and outside the flux tube and the last term being the buoyancy force.  Using Eq.~(\ref{apB_intot}) in Eq.~(\ref{apcurve}) we obtain:
\begin{eqnarray}
F_\perp & = & (a^2 - 1) ({\bf B}_{0}\cdot\grad {\bf B}_{0})\cdot { {\bf e}_\perp} + aB_0{\bf B}_{0}\cdot\grad \left( \frac{|{ {\bf e}_\perp}|^2}{B_0}c \right) \nonumber \\ &+&c{ {\bf e}_\perp}\cdot\grad \left( c \,{ {\bf e}_\perp} \right) \cdot{ {\bf e}_\perp} + ({\bar \rho}_0(\psi)e^{-\frac{m\phi(\psi, l)}{T_0(\psi)}}  - {\bar \rho}_0(\psi_0)e^{-\frac{m\phi(\psi, l)}{T_0(\psi_0)}}){ {\bf e}_\perp } \cdot \grad\phi.
\label{apnonlineq}
\end{eqnarray}
This equation determines the force from the shape of the field line, ${\tilde \psi}(\theta, \psi_0)$, for each $\psi_0$.  Note that by definition ${\bf B_{in}}\cdot\grad\psi_0 = (a\bb\cdot\grad + c{ {\bf e}_\perp}\cdot\grad)\psi_0 = 0$ so we can treat $\psi_0$ as a constant in Eq.~(\ref{apnonlineq}).  For an infinitesimal perturbation (${\tilde \psi} \ll \psi_0$) we obtain (expanding Eq.~(\ref{apnonlineq})) the "ballooning" equation:  
\begin{eqnarray}
F_\perp  \sim   B_0{\bf B}_{0}\cdot\grad \left( \frac{|{ {\bf e}_\perp}|^2}{B^2_0}{\bf B}_{0}\cdot\grad {\tilde \psi} \right) +    \frac{2\delta p}{B_0^2} ({\bf B}_{0}\cdot\grad {\bf B}_{0})\cdot { {\bf e}_\perp} + \delta\rho \;{ {\bf e}_\perp}\cdot\grad\phi
\label{aplineq}
\end{eqnarray}
where 
\begin{eqnarray}
\delta p = p_0(\psi, l) - p_{in}(\psi, l, \psi_0) = {\tilde \psi}\left(\frac{d{\bar p}_0}{d\psi} + \frac{1}{T_0}\frac{d{T}_0}{d\psi}\frac{m\phi}{T_0}{\bar p}_0\right)e^{-\frac{m\phi}{T_0}} 
\nonumber \\ \delta \rho = \rho_0(\psi, l) - \rho_{in}(\psi, l, \psi_0) = {\tilde \psi}\left(\frac{d{\bar \rho}_0}{d\psi} + \frac{1}{T_0}\frac{d{T}_0}{d\psi}\frac{m\phi}{T_0}{\bar \rho}_0\right)e^{-\frac{m\phi}{T_0}}  
\label{aplineq2}
\end{eqnarray}
and to linear order the difference between $\psi$ and $\psi_0$ in these expressions is irrelevant.  Also to linear order ${\tilde \psi} = \bxi\cdot\grad \psi_0$ so that we can write $\bxi = ({\tilde \psi}/B_0){ {\bf e}_\perp}$. The three terms in Eq.~(\ref{aplineq}) arise from three physical effects: the first is the extra bending of lines by the perturbation and is stabilising; the second is the change of the field line bending force due to the change of field strength (sometimes called the interchange drive); and the third is the gravitational/buoyancy force.  When gravity can be ignored ($\phi = constant$) Eq.~(\ref{aplineq}) reduces to the familiar ballooning equation of \cite{CHT}.

Like the simple straight line case the more general equilibria can be obtained from a variational principle.  Consider the magnetic energy functional:
\begin{eqnarray}
{\cal E_B}({\tilde\psi}(l,t), \psi_0)& = &\int {\bf B_{in}}\cdot d{\bf r} = \int {B_{in}}|d{\bf r}| \nonumber \\ 
\nonumber \\ & = & \int {B_{in}}\sqrt{\left(1+s\left(\frac{\partial {\tilde\psi}}{\partial l}\right)_{\psi_0}\right)^2 + w^2\left(\frac{\partial {\tilde\psi}}{\partial l}\right)_{\psi_0}^2}\frac{dl}{u}
\label{apEdef}
\end{eqnarray}
where the path of integration is along the perturbed field line -- i.e. at fixed $\psi_0$.  We shall assume that the integration terminates either in a boundary or at a distance where the field line is effectively unperturbed ({\em i.e.} $\bf B_0 = B_{in}$).  Now consider the variation of ${\cal E_B}$ due to an infinitesimal displacement of the field line of the form $\delta\bxi = \delta\xi { {\bf e}_\perp} = (\delta\psi/B_0){ {\bf e}_\perp}$ from the perturbed state keeping $\psi_0$ constant.  The change in the element of length is $\delta |d{\bf r}| = d{\bf r}\cdot\grad \delta\bxi \cdot {\bf B_{in}}/B_{in}$.  Thus:
\begin{eqnarray}
\delta{\cal E_B} & = &\int [{\delta B_{in}}|d{\bf r}| + { B_{in}}\delta |d{\bf r}|] = \int [\delta\bxi\cdot\grad_{\psi_0}{B}_{in}|d{\bf r}| + d{\bf r}\cdot\grad \delta\bxi \cdot {\bf B_{in}}] \nonumber \\ \nonumber \\ 
&= &- \int [{\bf B}_{in}\cdot\grad {\bf B}_{in} - \frac{1}{2}\grad_{\psi_0}{B}_{in}^2]  \cdot {\delta\xi {\bf e}_\perp } \frac{|d{\bf r}|}{B_{in}} = -\int F_\perp {\delta\xi} \frac{|d{\bf r}|}{B_{in}}
\label{apvariation}
\end{eqnarray}
where we have integrated by parts and used that $\delta\bxi\cdot {\bf B}_{in}$ must vanish at the boundaries or at the ends of the integration.  Clearly the equilibria $F_\perp =0$ are stationary points of ${\cal E_B}$.  It is also clear that motion against drag will push the field line to a minimum of ${\cal E_B}$.

\section{Flux Tube Boundary Layer}
\label{BoundaryLayer}
At the end of "line-tied" field lines (on the wall at $z=0$ and $z=L$) for the simple 
model equilibrium we set the Electric field in the $x$ and $y$ direction to zero -- {\em i.e.} $E_x(z=0, x, y, t)= E_x(z=L, x, y, t)= E_y(z=0, x, y, t) = E_y(z=L, x, y, t) = 0$. We also set the pressure and density to be unperturbed at the boundaries -- {\em i.e.} $p(z=0, x, y, t) = p(z=L, x, y, t) = p_0(x)$ and $\rho(z=0, x, y, t) = \rho(z=L, x, y, t) = \rho_0(x)$.  Motion along the field through the boundary  is not restricted.  However the flux tube solution we develop in Section~(\ref{fluxtube}) has  $B_z(z=0, x, y, t) = B_z(z=L, x, y, t) = B_0(x)\cos{\theta} \neq B_0(x)$ where $\theta (x, y)$ is the angle of the field line with the horizontal at the wall.  But since the field is line tied at the wall, the horizontal field leaving the wall must be $B_0$.   To rectify this problem a boundary layer of thickness $\Delta z \sim \delta_1$ forms at the wall.  In this appendix we examine the structure of this boundary layer at $z=0$.  To leading order the boundary layer solution does not affect the solution in the flux tube.

We take the field in the region $\Delta z \sim \delta_1$ to vary fast in both $z$ and $y$ -- {\em i.e.} $\frac{\partial}{\partial z}
\sim \frac{\partial}{\partial y} \sim \frac{1}{\delta_1}$.  The magnetic field is represented to lowest order (in $\delta_1/L$) by:
\begin{equation}
{\bf B} = \nabla\psi \times {\bf{\hat x}} + B_x(y,z){\bf{\hat x}}
\label{blfield}\end{equation}
We suppress labelling the slower variation in $x$ and the time dependence.  Note on the dynamical time of the boundary layer ($\Delta t = \delta_1 /V_A$) the time dependence is slow - thus to lowest order we seek an equilibrium.  Also note that the gravitational force is negligible to lowest order.  From parallel force balance ${\bf B}\cdot\nabla p = -\rho g {\bf B}\cdot {\bf{\hat x}}$ and $p = p_0(x)$ at $z=0$ we obtain the pressure variation in the boundary layer $\Delta p \sim \delta_1 \rho g$.  Then the lowest order boundary layer equilibrium is simply a force free magnetic equilibrium {\em i.e.} 
\begin{eqnarray}
{\bf J}\times{\bf B}= 0  \nonumber \\ 
\rightarrow B_x = B_x(\psi) \; and  \nonumber \\  \nabla^2 \psi = - \frac{1}{2}\frac{d}{d\psi}B_x^2
\label{bleq}\end{eqnarray}
At $z=0$, $B_z(y, z) = B_0$ therefore $\psi(y, 0) = -B_0y$.  As $z\rightarrow\infty$ the boundary layer solution must tend to the one dimensional solution $\psi \rightarrow \psi_0 (y)$ where $B^2 = B_x^2 + |\frac{d\psi_0}{dy}|^2 = B_0^2$.  The flux tube solution as $z\rightarrow 0$ is
\begin{equation}
B_x(y, 0) \equiv B_{xin}(y) = B_0\frac{(\frac{dx}{dz})}{(1 + (\frac{dx}{dz})^2)^{1/2}}
 \label{bxout}\end{equation}
where $\frac{dx}{dz}(y)= \tan{\theta}$ is evaluated from the flux tube solution at $z=0$.  Matching the boundary layer for $z\rightarrow\infty$ and flux tube solutions for $z\rightarrow 0$ gives:
\begin{equation}
 \psi_0 (y) = \int^y_0 dy' \sqrt{B_0^2 - B_{xin}^2(y')}. 
 \label{psi0}\end{equation}
The inverse of this function is $y = y_0(\psi_0)$.  Matching $B_x$ yields
\begin{equation}
 B_x(\psi) = B_{xin}(y= y_0(\psi)). 
 \label{Bxmatch}\end{equation}
 The boundary layer solution is then obtained from Eq.~(\ref{bleq}) with $B_x$ given by Eq.~(\ref{Bxmatch}) with
 the boundary conditions $\psi(y,0) = - B_0y$ and $\psi(y, z\rightarrow\infty) =  \psi_0 (y)$.
\end{appendix}
\end{document}